\documentclass{emulateapj}
\usepackage{natbib}
\usepackage{enumerate}
\usepackage{pdflscape}

\def\HI{\mbox{\ion{H}{1}}}
\def\HII{\mbox{\ion{H}{2}}}

\def\GII{\mbox{\ion{He}{2}}}

\def\ie{{\it i.e.}}
\def\eg{{\it e.g.}}
\def\ltsima{$\; \buildrel < \over \sim \;$}
\def\simlt{\lower.5ex\hbox{\ltsima}}
\def\gtsima{$\; \buildrel > \over \sim \;$}
\def\simgt{\lower.5ex\hbox{\gtsima}}

\begin{document}

\title{The First Galaxies and the likely discovery of their Fossils in the Local Group}
\author{Massimo Ricotti}
\affil{Department of Astronomy, University of Maryland, College Park,
  MD 20742}
\email{ricotti@astro.umd.edu}

\begin{abstract}
  In cold dark matter cosmologies, small mass halos outnumber larger
  mass halos at any redshift. However, the lower bound for the mass of
  a galaxy is unknown, as are the typical luminosity of the smallest
  galaxies and their numbers in the universe. The answers depend on
  the extent to which star formation in the first population of small
  mass halos may be suppressed by radiative feedback loops operating
  over cosmological distance scales. If early populations of dwarf
  galaxies did form in significant number, their relics should
  be found today in the Local Group. These relics have been termed
  ``fossils of the first galaxies''.  This paper is a review that
  summarizes our ongoing efforts to simulate and identify these
  fossils around the Milky Way and Andromeda.

  It is widely believed that reionization of the intergalactic medium
  would have stopped star formation in the fossils of the first
  galaxies. Thus, they should be among the oldest objects in the
  Universe. However, here we dispute this idea and discuss a physical
  mechanism whereby relatively recent episodes of gas accretion and
  star formation would be produced in some fossils of the first
  galaxies. We argue that fossils may be characterized either by a
  single old population of stars or by a bimodal star formation
  history.  We also propose that the same mechanism could turn small
  mass dark halos formed before reionization into gas-rich but
  starless ``dark galaxies''.

  We believe that current observational data supports the thesis that
  a fraction of the new ultra-faint dwarfs recently discovered in the
  Local Group are in fact fossils of the first galaxies.
\end{abstract}
  
\section{Introduction}

There are many questions that remain open in cosmology with regard to
the mass, number and properties of the smallest galaxies in the
universe. Have we already discovered the smallest galaxies in the
universe or we are still missing an elusive but large population of
ultra-faint dwarf galaxies?

In cold dark matter (CDM) cosmologies most of the dark halos that
formed before reionization had masses smaller than
$10^8-10^9$~M$_{\odot}$ \citep[\eg,][]{GnedinO:97}. The small mass
halos that survived tidal destruction to the modern epoch, were they
able to form stars, would constitute a sub-population of dwarf
satellites orbiting larger halos. Small mass dark halos significantly
outnumber more massive galaxies like the Milky Way and can be located
in the voids between luminous galaxies \citep[\eg,][]{Hoeft:06,
  Ricotti:09}.

However, until recently (\ie, before 2005) observations did not show a
large number of satellites around massive galaxies like the Milky Way
and Andromeda. This became known as the ``missing galactic satellite
problem, \citep{Klypin:99, Moore:99}. The voids between bright
galaxies appear to be devoid of dwarf galaxies \citep{Karachentsev:04,
  Karachentsevetal06, Tullyetal06}. While the abundance of dwarfs in
large voids may not pose a problem to CDM cosmology, as shown by
\cite{TinkerC:09}, it is unclear whether the predictions of the number
of faint dwarfs in the Local Group is consistent with both the number
of observed Milky Way dwarf satellites and the number of relatively
isolated dwarfs in the local voids.

Historically, the discrepancy between observation and theory on the
number of dwarf galaxies has been interpreted in two ways: 1) as a
problem with the CDM paradigm that could be solved by a modification
of the dark matter properties -- for instance by introducing warm dark
matter \citep[\eg,][]{BodeO:01} -- or 2) as an indicator of feedback
processes that are exceptionally efficient in preventing star
formation in small mass halos, which remain mostly dark
\citep[\eg,][]{HaimanAR:00}.

The recently discovered population of ultra-faint dwarfs
\citep{Belokurovetal06a, Belokurovetal07, Irwinetal07, Walshetal07,
  Willmanetal05ApJ, Willmanetal05AJ, Zuckeretal06a, Zuckeretal06b,
  Ibataetal07, Majewskietal07, Martinetal06} in combination with a
proper treatment of observational incompleteness \citep{SimonGeha07,
  Koposov:08, Tollerudetal08, Gehaetal08} has increased the estimated
number of Milky Way satellites to a level that can be more easily
reconciled with theoretical expectations. For instance, the
suppression of dwarf galaxy formation due to intergalactic medium
(IGM) reheating during reionization \citep{Babul:92, Efstathiou:92b,
  ShapiroG:94, HaimanRL:96, Thoul:96, Quinn:96, Weinberg:97,
  NavarroS:97, Bullock:00, Gnedin:00, Somerville:02, Dijkstra:04,
  Shapiro:04, Hoeft:06, OkamotoGT08, Ricotti:09}, in conjunction with
a strong suppression of star formation in small mass pre-reionization
dwarfs, may be sufficient to explain the observed number of Milky Way
satellites. In the near future we can hope to answer perhaps a more
interesting question: what is the minimum mass that a galaxy can have?
This is a non trivial and fundamental question in cosmology. Answering
it requires a better understanding of the feedback mechanisms that
regulate the formation of the first galaxies before reionization and
the details of the process of reionization feedback itself.

The formation of the first dwarf galaxies - before reionization - is
self-regulated on cosmological distance scales. This means that the
fate of small mass halos (\ie, whether they remain dark or form stars)
depends on local and global feedback effects. This type of galaxy
feedback differs from the more familiar model operating in normal
galaxies (\eg, SN explosions, AGN feedback, etc), where the feedback
is responsible for regulating the star formation rate within the
galaxy itself but does not impact star formation in other distant
galaxies. Rather, before reionization, each proto-dwarf galaxy reacts
to the existence of all the others.  Different theoretical assumptions
and models for the cosmological self-regulation mechanisms will, of
course, produce different predictions for the number and
luminosity of the first dwarf galaxies \citep{HaimanAR:00,
  RicottiGnedinShull02b, WiseA:08, OShea:08, RicottiGnedinShull08}.

We now introduce the basic concepts on how feedback-regulated
galaxy formation operates in the early universe (\ie, before
reionization):\\ A cooling mechanism for the gas is required in order to
initiate star formation in dark halos. In proto-galaxies that form after
reionization this is initially provided by hydrogen Lyman-alpha
emission. This cooling is efficient at gas temperatures of 20,000 K
but becomes negligible below $T \sim 10,000$~K. Later, as the
temperature drops below 10,000 K, the cooling is typically provided by
metal line cooling. In the first galaxies, however, both these cooling
mechanisms may be absent. This is because the first dwarf galaxies
differ when compared to present-day galaxies in two respects: 1) they
lack important coolants -- such as carbon and oxygen -- because the
gas is nearly primordial in composition, and 2) due to the smaller
typical masses of the first dark halos, the gas initially has a
temperature that is too low to cool by Lyman-alpha emission.

The gas in small mass halos with circular velocity $v_{\rm
  vir}=(GM_{\rm tot}/r_{\rm vir})^{1/2} \simlt 20$~km~s$^{-1}$, where
$r_{\rm vir}$ is the virial radius -- roughly corresponding to a mass
$M_{\rm tot} \simlt 10^8$ M$_{\odot}$ at the typical redshift of
virialization -- has a temperature at virialization $T \simlt 10,000$
K. Hence, if the gas has primordial composition it is unable to cool
by Lyman-alpha emission and initiate star formation unless it can form
a sufficient amount of primordial H$_2$ (an abundance $x_{H_2} \simgt
10^{-4}$ is required).  Because molecular hydrogen is easily destroyed
by far-ultraviolet (FUV) radiation in the Lyman-Werner bands
($11.3<h\nu<13.6$~eV) emitted by the first stars, it is widely
believed that the majority of galaxies with $v_{\rm
  vir}<20$~km~s$^{-1}$ remain dark
\citep[\eg,][]{HaimanAR:00}. However, several studies show that even
if the FUV radiation background is strong, a small amount of H$_2$ can
always form, particularly in relatively massive halos with virial
temperature of several thousands of degrees \citep{WiseA:07,
  OShea:08}. Thus, negative feedback from FUV radiation may only delay
star formation in the most massive pre-reionization dwarfs rather than
fully suppress it \citep{Machacek:00, Machacek:03}. We will argue
later that hydrogen ionizing radiation ($h\nu > 13.6$~eV) in the
extreme-ultraviolet (EUV) plays a far more important role in
regulating the formation of the first galaxies than FUV
radiation. Thus, in our opinion, models that do not include 3D
radiative transfer of H and He ionizing radiation cannot capture the
most relevant feedback mechanism that regulates galaxy formation in
the early universe \citep{RicottiGnedinShull02a,
  RicottiGnedinShull02b}.

After reionization, the formation of dwarf galaxies with $v_{\rm
  vir}<20$~km~s$^{-1}$ is strongly inhibited by the increase in the
Jeans mass in the IGM.  Thus, according to this model, reionization
feedback and negative feedback due to H$_2$ photo-dissociation by the
FUV background (important before reionization) determine the mass of
the smallest galactic building blocks.  The resultant circular
velocity of the smallest galactic building blocks is $v_{\rm vir} \sim
20$ km~s$^{-1}$, roughly corresponding to masses $M_{\rm tot} \sim
10^8-10^9$~M$_\odot$. If this is what really happens in the early
universe, the ``missing Galactic satellite problem'' can be considered
qualitatively solved because the predicted number of Milky Way
satellites with $v_{\rm vir}>20$ km~s$^{-1}$ is already comparable to
the estimated number of observed satellites after applying
completeness corrections (although this model may still have problems
reproducing the observations in detail).

\begin{figure*}[pth]
\epsscale{1.1}
\plotone{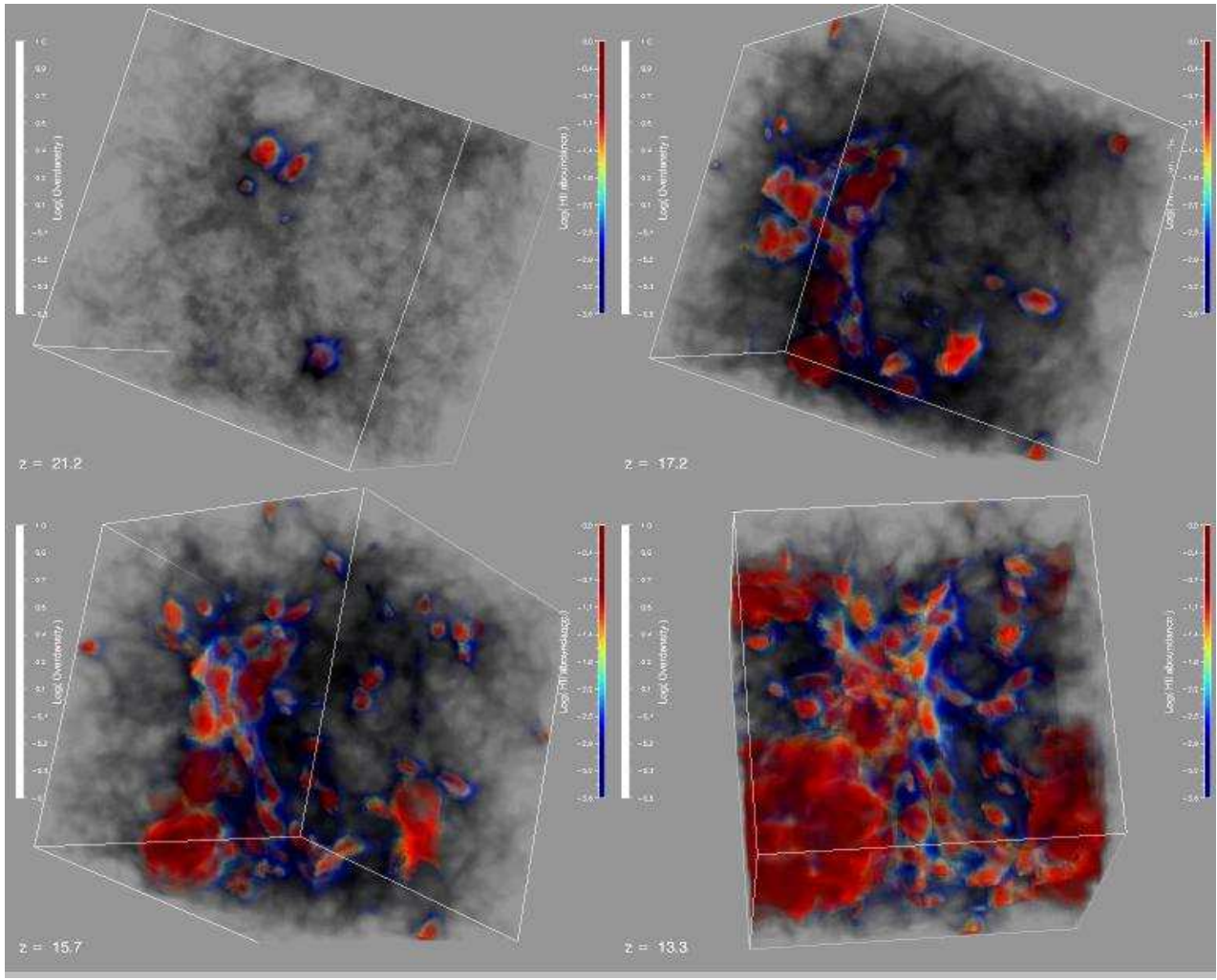}
\caption{3D rendering of cosmological \HII~ regions (fully ionized gas
  is red and partially ionized gas is blue) around the first galaxies
  in a box of 1.5~Mpc. The four boxes show a time sequence at
  redshifts $z=21.2, 17.2, 15.7, 13.3$ for the simulation S2 from
  \cite{RicottiGnedinShull02b}. The rendering shows several tens of
  small size \HII~ regions around the first galaxies (there are a few
  hundreds of galaxies in this volume). A movie of the same simulation
  shows that the \HII~ regions are short lived: they form, expand to a
  size comparable to the large scale filamentary structure of the IGM
  and recombine, promoting the formation of molecular hydrogen inside
  the relic \HII~ regions.}
   \label{fig:pfr2}
\end{figure*}

However, as briefly mentioned above, we have argued for some time that
most simulations of the first galaxies cannot capture the main
feedback mechanism operating in the early universe because they do not
include a key physical ingredient: radiative transfer of H and He
ionizing radiation. Our simulations of the formation of the first
galaxies are to date the only simulations of a cosmologically
representative volume of the universe (at $z \sim 10$) that include 3D
radiative transfer of H and He ionizing radiation
\citep{RicottiGnedinShull02a, RicottiGnedinShull02b,
  RicottiGS:08}. Figure~\ref{fig:pfr2} shows the evolution of ionized
bubbles around the first galaxies in a cubic volume of 1.5~Mpc in size
at redshifts $z=21.2, 17.2, 15.7, 13.3$ from one of our
simulations. The results suggest that negative feedback from the FUV
background is not the dominant feedback mechanism that regulates
galaxy formation before reionization. Rather, ``positive feedback'' on
H$_2$ formation from ionizing radiation \citep{HaimanRL:96,
  RicottiGS:01} dominates over the negative feedback of H$_2$
dissociating radiation. Hence, a strong suppression of galaxy
formation in halos with $v_{\rm vir}<20$~km~s$^{-1}$ does not take
place. In this latter case, some galactic satellites would be the
fossil remnants of the first galaxies. Comparisons of simulated
pre-reionization fossils to dwarf spheroidals in the Local Group show
remarkable agreement in properties \citep[][hereafter
RG05]{RicottiGnedin05}. Based on the results of the simulations, we
also suggested the existence of the ultra-faint population before it
was discovered about a year later \citep[see RG05,][]{BovillR:09}.

\subsection{Definition of ``pre-reionization fossils'' }

Throughout this paper we define ``pre-reionization fossils'' as the
dwarfs hosted in halos with a maximum circular velocity remaining
below 20~km~s$^{-1}$ at all times during their evolution: $v_{\rm
  max}(t)<20$~km~s$^{-1}$. It will become clear in this paper
that this definition is {\it not} directly related to the ability of
fossils to retain gas and form stars after reionization: in
\S~\ref{sec:infall} we describe a mechanism in which small mass halos
with $v_{\rm max}(t)< 20$~km~s$^{-1}$ are able to have a late phase of
gas accretion and possibly star formation.

Our definition of fossil reflects the {\it special cooling mechanisms
  and feedback processes} that regulate star formation and the number
of luminous halos with $v_{\rm max}(t)<20$~km~s$^{-1}$, before and
{\it after} reionization. In proto-fossil galaxies -- even adopting
the most conservative assumption of maximum efficiency of shock
heating of the gas during virialization -- the gas is heated to a
temperature below $T \sim 10,000$~K. Thus the gas cannot cool by
Lyman-alpha emission, a very efficient coolant. The cooling of the gas
is dominated either by H$_2$ roto-vibrational line emission or by
metal cooling, important if the metallicity exceeds $Z \sim
10^{-3}$~Z$_\odot$ \citep[\eg,][]{BrommF:01, SantoroS:06,
  RicottiGS:08, Smithetal:09}. These coolants are much less efficient
than Lyman-alpha emission. Moreover, H$_2$ abundance and cooling is
modulated and often suppressed by the FUV and EUV radiation
fields. The FUV radiation in the H$_2$ Lyman-Werner bands and hard
ultraviolet radiation have large mean free paths with respect to the
typical distances between galaxies, thus their feedback is global in
nature. Qualitatively, this explains why the first galaxies have low
luminosities and low surface-brightness, similar to dwarf spheroidal
(dSph) galaxies in the Local Group \citep[RG05,][]{BovillR:09,
  SalvadoriF:09}.

Simulations also show that stars in the the first galaxies do not form
in a disk but in a spheroid \citep[RG05,][]{RicottiGS:08}. A thin
galactic disk is not formed because of the high merger rates and the
low masses of dark halos in the early universe. Roughly,
pre-reionization fossils have a mass at virialization $M_{\rm
  tot}<10^8$~M$_\odot$, assuming they form at $z_{\rm vir} \sim 10$,
but their mass may increase by up to one order of magnitude by $z=0$
due to secondary infall \citep[RGS02a, b,][]{
  RicottiGnedinShull08}. Secondary infall does not affect $v_{\rm
  max}$, which remains roughly constant after virialization.

\subsection{Pre-reionization fossils and reionization}

The critical value of $v_{\rm max, crit}$ for which dwarf galaxy
formation is suppressed by reionization feedback is close to the
20~km~s$^{-1}$ value that defines a fossil, but it is {\it not necessarily
  the same value}. Indeed, it can be significantly larger than
20~km~s$^{-1}$ if the IGM is heated to $T \gg 10,000$~K
\citep{RicottiGS:00}. Thus, we expect that the virialization of new
``pre-reionization fossils'' is strongly suppressed after reionization
due to IGM reheating (\ie, they mostly form before
reionization). However, pre-reionization fossils and dark halos with
$v_{\rm max}<20$~km~s$^{-1}$ that virialized before reionization may
accrete gas and, in certain cases, form new stars after reionization at
redshifts $z<1-2$ \citep{Ricotti:09}.

Unfortunately, the value of $v_{\rm max, crit}$ is uncertain due to
our poor understanding of the thermal history of the IGM
\citep{RicottiGS:00}. The uncertainty surrounding the IGM equation of
state may partially explain the differences found in literature on the
values for $v_{\rm max, crit}$ and the different levels of suppression
of star formation as a function of the halo mass after reionization
\citep[\eg,][]{Weinberg:97, Gnedin:00, Hoeft:06,
  OkamotoGT08}. Regardless of assumptions for the reionization
feedback model, one should bear in mind that no halo with $v_{\rm max}
<20$~km~s$^{-1}$ can form stars after reionization unless the gas in
those halos has been significantly pre-enriched with metals. For
instance, the model by \cite{Koposov:09} assumes star formation after
reionization in halos as small as $v_{\rm max}\sim
10$~km~s$^{-1}$. With this assumption they find that their model is
consistent with observations of ultra-faint dwarfs but claim that
fossils are not needed to explain the data. However, star formation in
such small halos can only take place in a gas that was pre-enriched
with metals, suggesting the existence of older populations of stars in
those halos. Indeed, according to our definition, the smallest
post-reionization dwarfs with $10~{\rm km~s}^{-1}< v_{\rm max} <
20$~km~s$^{-1}$ in the \cite{Koposov:09} model are ``fossils''. As
stated above, fossils may also be able to form stars after
reionization due to a late phase of cold gas accretion from the IGM
\citep{Ricotti:09}.

\subsection{Identification of pre-reionization fossils in observations}

Pre-reionization fossils are not easily identifiable because $v_{\rm
  max}$ cannot be measured directly from observations. Understanding
the star formation history of dwarf galaxies may help in this respect,
as fossils likely show some degree of suppression of their star
formation rate occurring about 12.5 Gyr ago due to
reionization. However, their identification based on their star
formation history may be complicated because some pre-reionization
fossils in the last 10~Gyr may have had a late phase of gas accretion
and star formation. The caveat is that star formation histories cannot
be measured with accuracy better than to within 1-2~Gyr and the
accuracy becomes increasingly poorer for old stellar
populations. Thus, it is impossible to prove whether an old population
of stars formed before reionization (which happened about 1~Gyr after
the Big Bang) or at $z \sim 3$, when the Milky Way was
assembled. Nevertheless, ultra-faint dwarfs that show some degree of
bimodality in their star formation history are candidates for being
pre-reionization fossils.

According to results by RG05 and \cite{BovillR:09}, Willman~1,
Bootes~II, Segue~1 and Segue~2 do not lie on the luminosity-surface
brightness relationship of simulated pre-reionization fossils. This
result is based on the assumption that fossil properties are not
modified by tides.  Their surface brightness is larger than the model
predictions for objects with such low-luminosity. An as yet
undiscovered population of ultra-faints with lower surface brightness
is instead predicted by our simulations. It is likely that the
properties of the lowest luminosity ultra-faints may have been
modified by tidal forces due to their proximity to the Milky Way disk.

Although it is difficult to identify individual fossils, statistical
arguments suggest that at least some ultra-faint dwarf galaxies are
pre-reionization fossils. This is because the number of satellites
from N-body simulations with $v_{\rm max}(t)>20$~km~s$^{-1}$ is
substantially smaller than the estimated number of observed satellites
after completeness corrections. Admittedly the current theoretical and
observational uncertainties on the number of satellites are still
large. However, if the estimated number (after completeness
corrections) of ultra-faint dwarfs increases further, the existence of
pre-reionization fossils will be inescapably proven. This is
especially the case if a population of ultra-faint dwarfs with
luminosities similar to Willman~1, Bootes~II, Segue~1 and Segue~2 but
surface brightness below the current sensitivity limit of the SDSS --
as predicted by our simulations -- is discovered.
 
The possibility of identifying the fossils of the first galaxies in
our own backyard is very exciting. It would greatly improve our
understanding of the physics involved in self-regulating the formation
of the first galaxies before reionization. Clearly, even the launch of
the James Webb Space Telescope ({\it JWST}), would not yield the
wealth of observational data on the formation of the first galaxies
that could be obtained by studying ultra-faint galaxies in the Local
Group.

The rest of the paper is organized as follows. In \S~\ref{sec:obs} we
briefly review and discuss observational data on Galactic satellites,
in \S~\ref{sec:theo} we summarize the results of simulations of the
formation of the first galaxies in a cosmological volume and the
effect of reionization feedback on galaxy formation. In
\S~\ref{sec:infall} we discuss a recently proposed model for ``late
gas accretion'' from the IGM onto small mass halos. In
\S~\ref{sec:comp} we compare the theoretical properties of simulated
pre-reionization fossils to observations. In \S~\ref{sec:disc} we
compare different ideas for the origin of classical and ultra-faint
dwarf spheroidals. We present our conclusions in \S~\ref{sec:conc}.

\section{Observations}\label{sec:obs}

\subsection{The ultra-faint satellites of the Milky Way and Andromeda}

Prior to 2005, the number of observed dwarf satellites of the Milky
Way and Andromeda was about 30 \citep{Mateo98}. One of the most
evident properties of the dwarfs in the Local Group is a type
segregation, with ``gas free'' dwarf spheroidal (dSph) galaxies
distributed near the center of their host galaxy and gas rich dwarf
Irregulars (dIrr) at larger distances from the galactic
centers. Notable exceptions are the Magellanic Clouds that are dIrr
less than 100~kpc from the center of the Milky Way and a few isolated
dSphs like Tucana and Antlia. One popular explanation for this
segregation is the transformation of dIrr into dSph due to tidal and
ram pressure stripping as dwarfs fall toward the Milky Way center
\citep{Mayeretal07, PenarrubiaNM08}. In addition, simulations showed
that the number of dark matter satellites of the Milky Way with mass
$>10^8$~M$_\odot$ (\ie, with mass sufficiently large to expect star
formation in them) was an order of magnitude larger than the number of
known dwarf satellites \citep{Klypin:99, Moore:99}. This posed a
problem for CDM cosmologies.

Since 2005-2006 the number of known Local Group satellites has begun
to increase dramatically, with the discovery of a new population of
ultra-faint dwarfs.  The new galaxies have been discovered by data
mining the SDSS and other surveys of the halo around M31, resulting in
the discovery of 14 new ultra-faint Milky Way satellites
\citep{Belokurovetal06a, Belokurovetal07, Irwinetal07, Walshetal07,
  Willmanetal05ApJ, Willmanetal05AJ, Zuckeretal06a, Zuckeretal06b,
  Koposov:08, Walsh:09} and 11 new companions for M31
\citep{Ibataetal07, Majewskietal07, Martinetal06, Martinetal09}.
Unofficial reports from members of the SDSS collaboration state that
there are actually at least 17 new ultra-faint Milky Way dwarfs, but
several of them are as yet unpublished (anonymous referee's private
communication).

The new Milky Way satellites have been slowly discovered since SDSS
Data Release 2, with the most recent, Segue 2 discovered in Data
release 7 \citep{Adelman-McCarthyetal06,Adelman-McCarthyetal07}.
\cite{Koposov:08} and \cite{Walsh:09} systematically searched Data
Releases 5 and 6, respectively. Due to the partial sky coverage of the
SDSS, and assuming isotropic distribution of satellites \citep[but
see][]{Kroupa:05, Zentner:05}, the total number of ultra-faint dwarfs
in the Milky Way should be at least 5.15 times larger than the
observed number \citep{Tollerudetal08}. With this simple but
conservative correction, the number of Milky Way satellites within
$400$~kpc is about $12 + 5.15 \times (14 \pm 3.7) \sim 84 \pm
19$. The quoted uncertainty is simply Poisson error due to the
relatively small number of known ultra-faint dwarfs.

In estimating the completeness correction for the number of Milky Way
dwarfs, one should account for selection effects inherent in the
method used to find the ultra-faints in the SDSS data. In addition to
completeness corrections for the survey's coverage of the sky, the
most important selection effect is the total number of stars from the
object seen in the survey, \ie, horizontal branch stars or main
sequence and/or red giant stars for the lowest luminosity ultra-faints
like Coma or Segue~1. This sets a limiting surface brightness cutoff
at roughly 30~${\rm mag~arcsec}^{-2}$ for the SDSS \citep{Koposov:08}
(but see \cite{Martin:08} that find a limiting surface brightness
about 6.4 times larger). There is also a distance-dependent absolute
magnitude cutoff. The efficiency of finding ultra-faint dwarfs by data
mining the SDSS typically drops rapidly at Galactocentric distances
beyond $50-150$~kpc for the ultra-faints (depending on their
luminosity) \citep{SimonGeha07, Koposov:08, Tollerudetal08, Walsh:09}.
Of the new Milky Way dwarfs, only Leo T is well beyond this distance
threshold\footnote{Leo~T was found because it contains a young stellar
  population and gas. Otherwise, it would not have been identified as
  an ultra-faint dwarf due to its large Galactocentric distance.}, and
11 of the 14 new Milky Way satellites are within $200$~kpc.

The luminous satellites can be radially biased. So, the abundance of
the faintest satellites within 50~kpc, that is the most complete
sample, may not be used to correct for incompleteness at larger
distances from the Galactic center without prior knowledge of this
bias. And, of course, satellites of different luminosity and surface
brightness will have different completeness limits.  These selection
biases have been considered in a paper by \cite{Tollerudetal08}.
Their study finds that there may be between 300 to 600 luminous
satellites within $400$~kpc.  Their estimate for the number of
luminous satellites within a Galactocentric distance of $200$~kpc is
between 176 to 330.

Recent surveys of M31 \citep{Martinetal06, Ibataetal07, Martinetal09}
have covered approximately a quarter of the space around the M31
spiral.  The surveys have found 11 new M31 satellites, bringing the
total number to 19. If we make a simple correction for the covered
area of the survey, the estimated number of M31 satellites, including
the new dwarfs, increases from $8$ to $52 \pm 13$.

The sensitivity limits of the surveys for Andromeda do not allow the
detection of ultra-faint dwarfs that would correspond with those with
the lowest luminosity found in the Milky Way. However, despite the
fact that Andromeda and the Milky Way are thought to have
approximately the same mass (within a factor of two), their satellite
systems show interesting differences for even the satellites at the
brighter end of the luminosity function. For instance, there are small
differences in the galactocentric distance distribution of satellites
and in the morphology of the satellites (\eg, number of dIrr, dE and
dSphs).

\subsection{Peculiar ultra-faint dwarfs}\label{sec:leoT}

Many of the newly discovered dwarfs are dSphs with a dominant old
population of stars and virtually no gas, which makes them candidates
for being pre-reionization fossils. However, there are notable
exceptions that we discuss below that may not perfectly fit the
properties of simulated ``fossils''. For instance, the dwarf galaxy
Leo~T resembles all the other ultra-faints but has gas and recent star
formation \citep{Irwinetal07, SimonGeha07}. We argue that Leo~T
could be a true ``fossil'' with $v_{\rm max}(t)<20$~km~s$^{-1}$, but
may have experienced a late phase of gas accretion and star formation due
to the mechanism discussed in \S~\ref{sec:infall}.

{\it Leo~T} has a stellar velocity dispersion of $\sigma_{Leo T} = 7.5
\pm 1.6 {\rm km~s}^{-1}$ \citep{SimonGeha07}, or an estimated
dynamical mass of $10^7$~M$_\odot$ within the stellar spheroid
(although its total halo mass may be much larger).  Leo~T shows no
sign of recent tidal destruction by either the Milky Way or M31
\citep{deJongetal08} and is located in the outskirts of the Milky Way
at a Galactocentric distance of $400$~kpc.  Leo~T photometric
properties are consistent with those of pre-reionization fossils. On
the other hand, the halo of Leo~T could be sufficiently massive
($v_{\rm max}(t) >20$~km~s$^{-1}$) to retain or accrete mass after
reionization and not be a pre-reionization fossil. However, as
discussed in \S~\ref{sec:infall} it is also possible that Leo~T is a
pre-reionization fossil that has been able to accrete gas from the IGM
at late times due to an increase in the concentration of its dark halo
and a decrease in the IGM temperature \citep{Ricotti:09}.  Under this
scenario, Leo~T stopped forming stars after reionization, but was able
to start accreting gas again from the IGM at $z \simlt 1-2$ and
therefore would have a bimodal stellar population. \cite{deJongetal08}
have found evidence for bimodal star formation in Leo~T. Our model
would explain why Leo~T does not resemble other dIrr and is similar to
dSphs and pre-reionization fossils, while not suffering significant
tidal stripping.

{\it Willman~1, Bootes~II, Segue1 and Segue2} are among the lowest
luminosity ultra-faint dwarfs discovered so far; however, they do not
fit the typical properties of pre-reionization fossils (see
\S~\ref{sec:comp}).  For instance Willman~1 has a dynamical mass
within the largest stellar orbit ($r \sim 100$~pc) of $5 \times 10^{5}
M_{\odot}$ and a mass-to-light ratio $\sim 470$, similar to other
ultra-faint dwarfs \citep{Willmanetal05AJ}. However, given its low
luminosity, Willman~1 has central surface brightness that is too large
when compared to simulated ``fossils''. Simulated fossils with
luminosities $L_V<10^3$~L$_\odot$ should have a typical surface
brightness that falls below the detection limit of $\sim$30 ${\rm
  mag~arcsec}^{-2}$ estimated for the SDSS \citep{Koposov:08}. Hence,
the lowest luminosity fossils may still be undiscovered. Although the
nature of the lowest luminosity ultra-faints is unknown, due to
selection effects they can only be found within $\sim 50$~kpc of the
Galactic center. Thus it is possible, and perhaps to be expected, that
their properties have been affected by tidal forces
\citep{PenarrubiaNM08, MartinJR:08}.

\section{Formation of first galaxies in CDM}\label{sec:theo}
\begin{figure*}[thp]
%\epsscale{1.2}
\centering
\includegraphics[scale=0.9,angle=0]{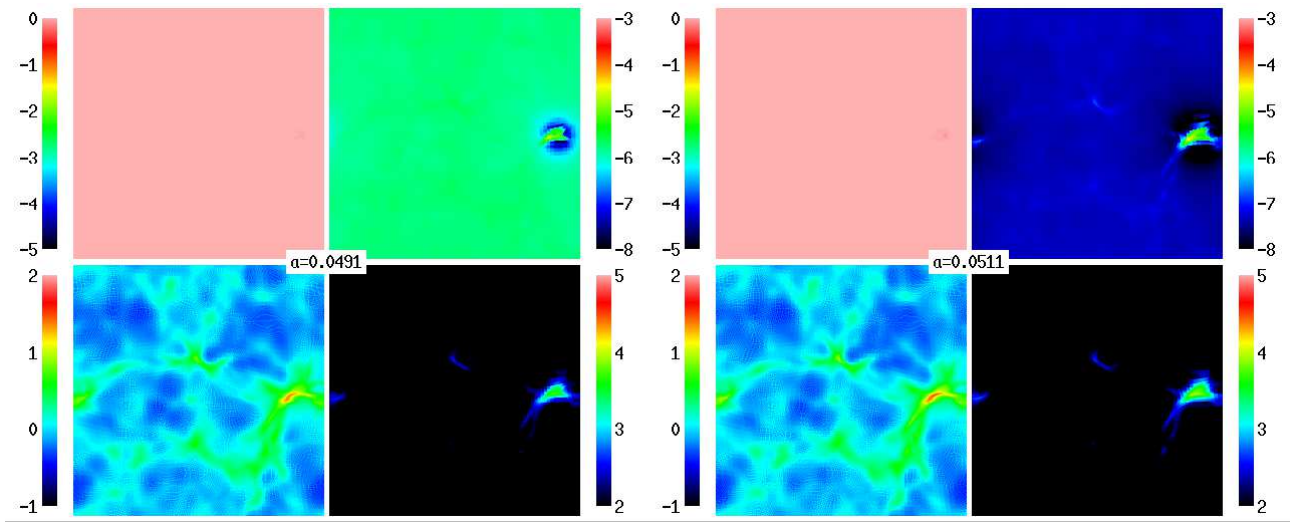}
\caption{The two (2x2) panels show slices through the most massive
  object in the simulation 64L05p2 in RGS02b at $z=19.4$ and
  $18.5$. The box size is $L_{box}=0.7$ comoving Mpc.  Each one of the
  2x2 panels shows in log-scale: the neutral hydrogen fraction (top
  left), the molecular fraction (top right), the gas overdensity
  (bottom left), and the gas temperature (bottom right). The sequence
  illustrates the evolution of a H$_2$ dissociation sphere around a
  single source (panel at $z=19.4$) and the dominance of the H$_2$
  dissociating background at $z=18.5$.}
   \label{fig:pfr0}
\end{figure*}

The first episodes of star formation in the universe are thought to
take place at redshift $z \sim 30-50$, in the center of dark matter
halos with typical mass $M_{\rm tot} \sim 10^5-10^6$~M$_\odot$. The
gas in these halos is metal free and simulations show that a single or
binary massive star per halo is formed \citep{Bromm:02, Abel:02,
  Saigo:04, Gao:05, OShea:07, Yoshida:08, Turk:09, StacyGB:09}. Such
stars are called Pop~III and their mass, although not well constrained
by the simulations, is quite large: in the range between
$20$~M$_\odot$ to a few $100$~M$_\odot$. Whether or not we can refer
to minihalos containing a single massive star (or a binary star) in
their center as the ``first galaxies'' is debatable. However, the
crucial point to be made here is that there is a gap of 2 to 3 orders
of magnitude between the typical halo mass in which Pop~III stars
are born ($10^5-10^6$~M$_\odot$) and the typical mass of the
population of dwarf galaxies that are not strongly influenced by
radiative and reionization feedback ($10^8-10^9$~M$_\odot$). The
primordial dwarfs that fill the gap are those that we refer to as {\it
  pre-reionization fossils}.

If the formation of pre-reionization fossils is not drastically
suppressed by radiative feedback, their number may be several orders
of magnitude larger than the number of more massive dwarfs. This is
because in CDM cosmologies the number of dark halos per unit comoving
volume roughly scales with the mass as $N \propto M_{\rm dm}^{-2}$.

It is widely believed that nearly all halos with mass $M_{\rm
  tot}>10^8-10^9$~M$_\odot$ host luminous galaxies, although there can
be substantial disagreement among theorists on their
luminosity. However, most of the theoretical controversy rests in
understanding the fate of the halos with mass between
$10^6-10^8$~M$_\odot$ and the dominant feedback that determines
whether they become luminous or remain dark. We will elaborate on this
statement in the next sections.
\begin{figure*}[thb]
%\epsscale{1.2}
\centering
\includegraphics[scale=0.9,angle=0]{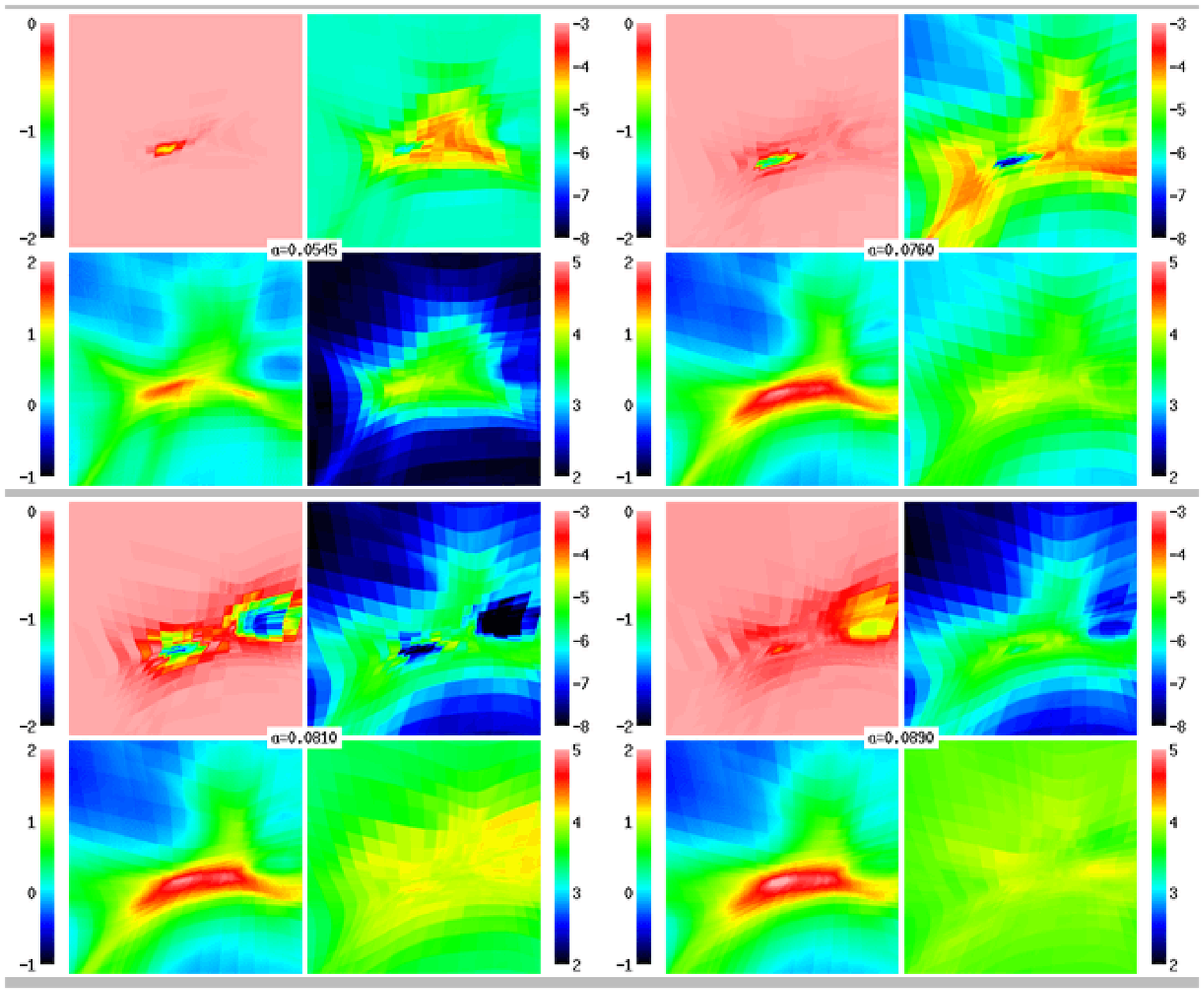}
\caption{Same as in Fig.~\ref{fig:pfr0} except for a zoomed region of
  $0.125^2$ $h^{-2}$Mpc$^2$ around the most massive object in the
  64L05p3 simulation in RGS02b. In this time sequence of images (top: $z=17.3,
  12.2$ from left to right; bottom: $z= 11.3, 10.2$ from left to
  right) we recognize the two main processes that create H$_2$ in the
  filaments: ``positive feedback regions'' in front of \HII~ regions
  and the reformation of H$_2$ inside relic \HII~ regions. The bursting
  mode of the star formation is evident from the continuous formation
  and recombination of the \HII~ regions in the time sequence of the
  slices.}
   \label{fig:pfr1}
\end{figure*}

\subsection{Radiative feedback}\label{sec:sims}

Simulating the formation of the first stars is a relatively well
defined initial condition problem, given the cosmological
parameters. However, these simple initial conditions must soon
be modified to take into account the effects of other newborn stars,
whose properties are still quite uncertain. The physics becomes more
complex as competing feedback effects determine the fate of the first
galaxies: radiative feedback regulates the formation and destruction
of H$_2$ and metals are injected into the IGM and into protogalaxies.

\subsubsection{Negative feedback from H$_2$ photo-dissociating radiation}

The net effect of radiative feedback on the global star formation
history of the universe before the redshift of reionization is
uncertain.  An FUV background (at energies between $11.34$~eV and
$13.6$~eV) destroys H$_2$, the primary coolant at the start of galaxy
formation. The FUV radiation emitted by the first few Pop~III stars is
sufficient to {\it suppress or delay} galaxy formation in halos with
circular velocities $v_{\rm vir}< 20$~km~s$^{-1}$ that are too small
to cool by Lyman-alpha emission
\citep{Haimanetal00,Ciardietal00,Machaceketal00, OShea:08}.  Hence,
according to this scenario, most halos with masses $<10^8-10^9$
~M$_\odot$ remain dark. More work is needed to quantify the level of
suppression of galaxy formation and examine how these models compare
to observations of Milky Way satellites.

Figure~\ref{fig:pfr0} illustrates the effect of H$_2$ dissociating
radiation on the IGM.  The two panels show slices through a simulation
in \cite{RicottiGnedinShull02b} at $z=19.44$ and $z=18.5$. The
top-right tiles in the two panels show H$_2$ abundance. At $z=19.4$,
the H$_2$ has its relic abundance everywhere in the IGM except inside
the dissociation spheres around the first galaxies, where it is
destroyed. At $z=18.5$, the dissociation spheres are still visible,
but the UV background starts to dissociate H$_2$ everywhere in the IGM
except the denser filaments.

\subsubsection{Positive Feedback Regions}

Our main criticism for the ``negative feedback'' model is that it does
not take into account the effect of hydrogen ionizing radiation
\citep{RicottiGnedinShull01, Whalenetal07} that, according to
simulations, may indeed play a dominant role in regulating galaxy
formation before reionization \citep{RicottiGnedinShull02a,
  RicottiGnedinShull02b}. Simulations including 3D radiation transfer
show that star formation in the first small mass halos is inefficient,
partially due to winds produced by internal UV sources.  This produces
galaxies that are extremely faint and have very low surface
brightnesses.  However, our simulations show that a large number of
ultra-faint dwarfs (a few hundred galaxies per comoving Mpc$^3$) form
before reionization at $z \sim 7-10$. Hence, according to this model,
the Local Group may contain thousands of ultra-faint dwarf galaxies.

Ionizing radiation from the first stars enhances the production of
H$_2$ (we refer to this as ``positive feedback'') by creating free
electrons and promoting the formation of H$^-$, the main catalyst for
the formation of H$_2$ in a low metallicity gas \citep*{ShapiroKang87,
  HaimanReesLoeb96, RicottiGnedinShull01, AlvarezBrommShapiro06,
  Ciardietal06}.  \cite{RicottiGnedinShull01} found that shells of
H$_2$ can be continuously created in precursors around the Str\"omgren
spheres produced by ionizing sources and, for a bursting mode of star
formation, inside recombining \HII~regions. We refer to these shells
as ``positive feedback regions''.  This is because the catalyst H$^-$,
and hence H$_2$, is formed most efficiently in regions where the gas
ionization fraction is about $50\%$. This local ``positive feedback''
is difficult to incorporate into cosmological simulations because the
implementation of spatially inhomogeneous, time-dependent radiative
transfer is computationally expensive.

Fig.~\ref{fig:pfr1} shows ``positive feedback regions'' in one of our
simulations. The figure shows a slice through a simulation at 4
different times (at $z= 17.3, 12.2, 11.3$, and 10.2). We recognize the
two main processes that create H$_2$ in the filaments. In the top-left
frame at $z=17.3$ we can see a ``positive feedback region'' as
an irregular shell of H$_2$ surrounding the \HII~region that is barely
intersected by the slice. In the bottom-left frame ($z=11.3$) two
\HII~ regions are clearly visible.  Inside the \HII~regions, the H$_2$
is destroyed. In the bottom-right frame ($z=10.2$) the \HII~regions
are recombining (demonstrating that the star formation is bursting)
and new H$_2$ is being reformed inside the relic \HII~regions. A finer
inspection\footnote{Movies of 2D slices and 3D rendering of the
  simulations are publicly available on the web at the URL:
  http://astro.umd.edu/$\sim$ricotti/movies.html} of the time
evolution of this slice shows that at least five \HII~regions form and
recombine between $z = 20$ and $z=10$ in this small region of the
simulation.

There are two reasons why our results are still controversial. First,
our simulations do not yet have sufficient resolution to ensure their
convergence. Second, there are no other published simulations to
compare our results with.  Only recently have some groups started to
include the effect of 3D radiative transfer on hydrodynamics
\citep[\eg, ][]{WiseA:07, WiseA:08}. However, currently there are no
other simulations of the formation of the first galaxies in a
cosmological volume suited for comparison with observations of dwarfs
in the Local Group other than our own
\citep{RicottiGnedinShull02a,RicottiGnedinShull02b, RicottiGnedin05,
  RicottiGnedinShull08, BovillR:09}. Hence, our results may differ
from other numerical studies because of the inclusion in the code of
the effects of ``positive feedback regions'' and galactic winds from
ionizing radiation.

Simulations by \cite{WiseA:08} include a self-consistent treatment of
hydro and 3D radiative transfer that is more accurate than our
approximate, but faster method. However, because the authors use
ray-tracing for the radiative transfer, only a few sources of
radiation can be simulated at the same time. This limits the volume
and number of galaxies that can be simulated. Due to these limitations
the simulations are not suited for comparison between the primordial
dwarf populations and the ultra-faint dwarfs. In addition, at the
moment, the aforementioned simulations do not include metal cooling
and the formation of normal stars (other than Pop~III).

\subsubsection{The simulations}

The simulation used for comparison to observations of ultra-faint
dwarfs has been thoroughly described in Ricotti, Gnedin \& Shull\
(2002a, 2002b) as run ``256L1p3''. Here we remind the reader that the
simulation includes $256^3$ dark matter particles, an equal number of
baryonic cells, and more than 700,000 stellar particles in a box of
size $\sim 1.5$~comoving Mpc. The mass of the dark matter particles in
our simulation is 4930~M$_\odot$, and real comoving spatial resolution
(twice the Plummer softening length) is $150~h^{-1}$~pc (which
corresponds to a physical scale of 24 parsecs at $z=8.3$). This
resolution allows us to resolve cores of all simulated galaxies that
would correspond to the observed Local Group dwarfs.  The stellar
masses are always smaller than the initial baryon mass in each cell
but can vary from $\sim 0.6$~h$^{-1}$~M$_\odot$ to
$600$~h$^{-1}$~M$_\odot$ with a mean of
$6$~h$^{-1}$~M$_\odot$. Stellar particles do not represent individual
stars but, in general, a collection of stars (\eg, OB associations).

The simulation includes most of the relevant physics, including
time-dependent spatially-variable radiative transfer using the OTVET
approximation \citep{GnedinA:01}, detailed radiative transfer in
Lyman-Werner bands, non-equilibrium ionization balance, etc.  In
addition to primordial chemistry and 3D radiative transfer, the
simulations include a sub-grid recipe for star formation, metal
production by SNe and metal cooling. The code also includes mechanical
feedback by SN explosions. However, we found that for a Salpeter IMF,
the effect of SNe is not dominant when compared to feedback produced
by ionizing radiation from massive stars \citep[][hereafter
RGS08]{RicottiGnedinShull08}. The effect of SN explosions is somewhat
model dependent and uncertain because it is treated using a sub-grid
recipe. Hence, the simulation analyzed in this work includes metal
pollution but not mechanical feedback by SNe.

In RG05, we included the effect of reionization in the simulation
256L1p3. Because the size of the simulation box has been fixed at
$\sim 1.5$ comoving Mpc, the simulation volume is too small to model
the process of cosmological reionization with sufficient accuracy. We
therefore assume that the simulation volume is located inside an \HII~
region of a bright galaxy at a higher redshift. Specifically, we
introduce a source of ionizing radiation within the computational box,
properly biased, which corresponds to a star-forming galaxy with the
constant star formation rate of 1 solar mass per year (similar to star
formation rates of observed Lyman Break Galaxies at $z \sim 4$,
Steidel et al.\ 1999). The source is switched on at $z=9.0$, and by
$z=8.3$ the whole simulation box is completely ionized.
\begin{figure*}[th]
\plottwo{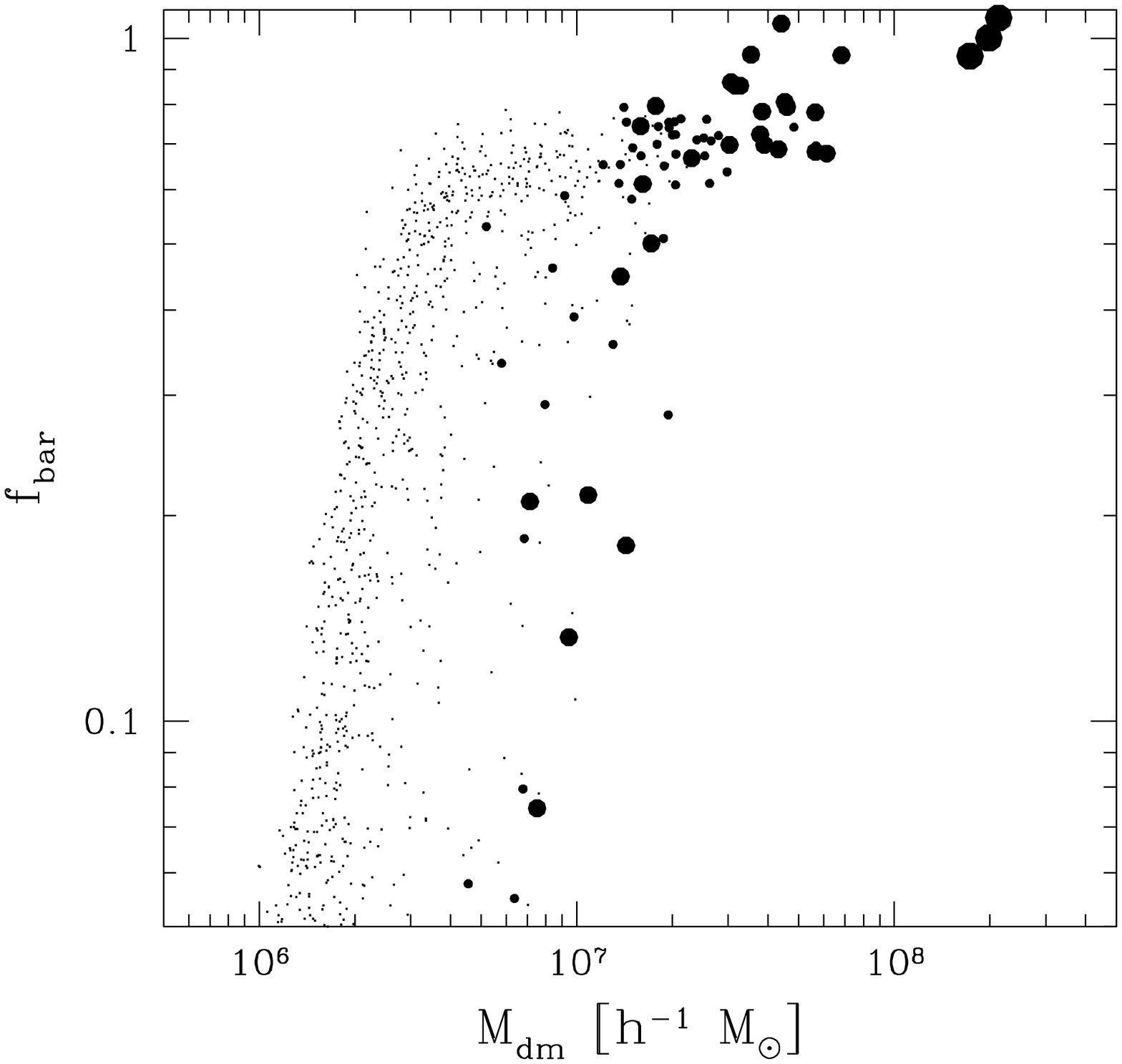}{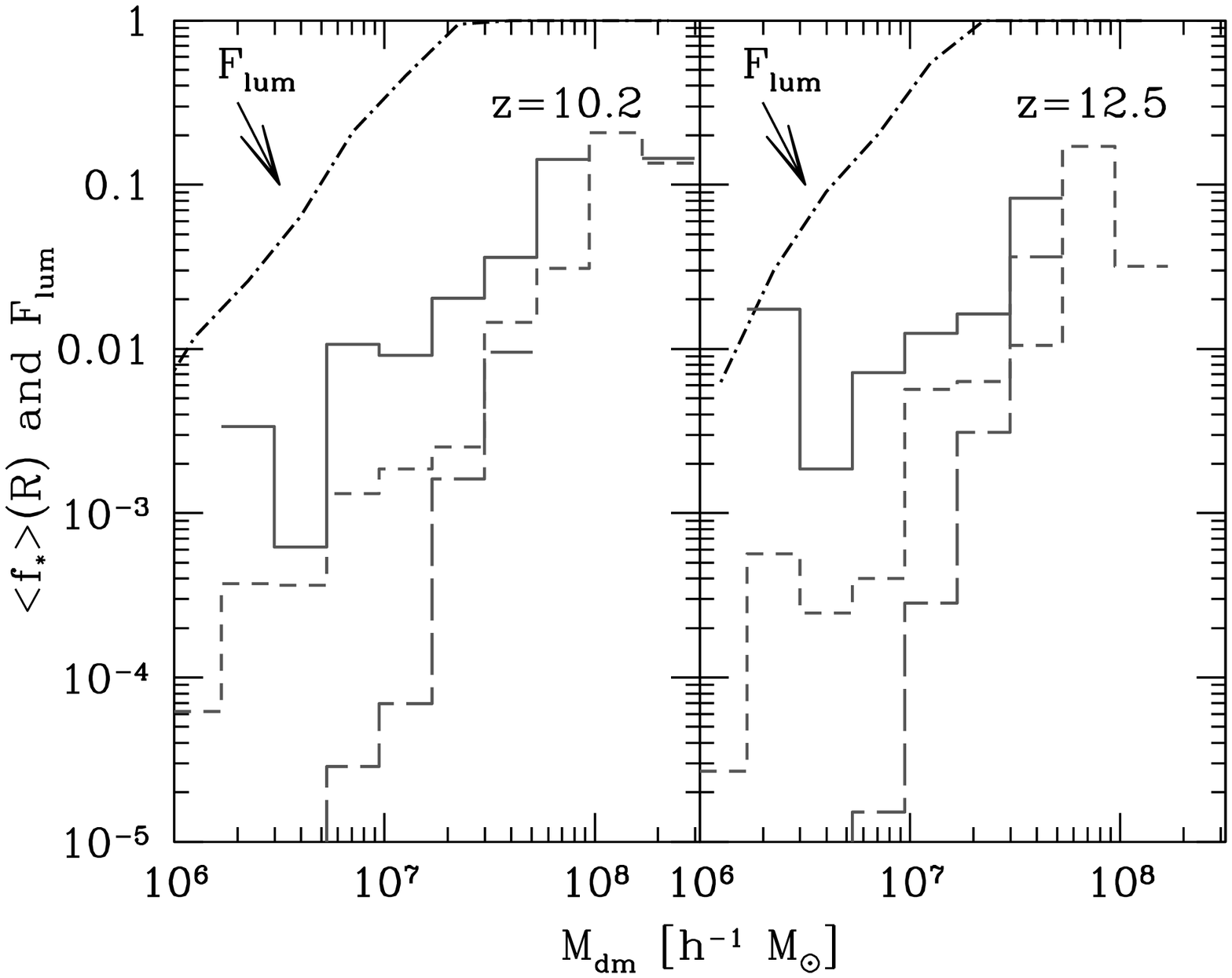}
\caption{{\it (Left)}. Fraction of baryons retained by each galaxy
  (normalized to the cosmic mean value, $M_{\rm bar}=f_{\rm bar}^{\rm
    max}M_{\rm dm}$, where $f_{\rm bar}^{\rm max}=
  \Omega_b/\Omega_m=0.136$) as function of the halo mass $M_{\rm dm}$
  for run S1 in RGS08 at $z=10$ . The size of the dots is proportional
  to the fraction of stars $f_*=M_*/M_{\rm bar}$ in each halo: from
  the largest to the smallest dots we have $f_*>10$\%, $1<f_*<10$\%,
  $0.1<f_*<1$\% and $f_*<0.1$\%, respectively. The plot illustrates
  the role of internal and external sources of ionizing radiation in
  reducing the gas retained by small mass halos. Luminous sources
  (with larger $f_*$) retain less gas than dark halos due to winds
  driven by internal sources of radiation. {\it (Right)}. Average star
  formation efficiency $\langle f_* \rangle$ as a function of halo
  mass at $z=10.2$ (left panel) and $12.5$ (right panel) for run
  S1. We divide all halos into three groups: those at distance $d < 8$
  kpc from the nearest luminous halo (solid histograms), those with
  $8~{\rm kpc} < d < 50$~kpc (dashed histogram), and those with $d >
  50$ kpc (long-dashed histograms). The dot-dashed curve shows the
  fraction of luminous halos $F_{\rm lum}(M_{\rm dm})$ as a function
  of the halo mass.}
   \label{fig:feed}
\end{figure*}

\subsubsection{Summary of main results}

The main results of the simulations are the following \citep[see][for
details]{RicottiGnedinShull08}:
\begin{enumerate}
\item {\em Negative feedback}. H$_2$ photo-dissociation from FUV
  radiation, the main negative feedback thought to suppress the
  formation of the first galaxies, is not the dominant feedback. If we
  modify the spectrum of the sources of radiation to artificially
  increase or decrease the FUV flux emitted by up to one order of
  magnitude, we do not find any appreciable effect on the global star
  formation history.
\item {\em Self-regulation}. Feedback by hydrogen ionizing radiation
  (EUV) plays the key role. Galactic outflows, produced by UV
  photo-heating from massive stars, and H$_2$
  formation/photo-dissociation induces a bursting star formation mode
  in the first galaxies that acts as the catalyst for H$_2$
  re-formation inside relic (recombining) \HII~regions and in the
  ``precursors'' of cosmological Str\"omgren spheres ({\it i.e.},
  positive feedback regions).  As a result, the formation of the first
  galaxies is self-regulated on a cosmological distance scale. It is
  significantly reduced by radiative feedback but it is not completely
  suppressed, even in halos with $v_{\rm max} \sim
  5-10$~km~s$^{-1}$. Note that our sub-grid recipe for star formation
  assumes a Schmidt law with a tunable efficiency parameter
  $\epsilon_*$ (the fraction of gas converted into stars per crossing
  time). We find that the global star formation history and the
  fraction of baryons converted into stars in each galaxy,
  $f_*=M_*/M_{\rm bar}$, is nearly independent on the assumed value of
  $\epsilon_*$. This is typical for feedback regulated star
  formation. Hence, the star formation efficiency, $f_*$, is not an
  assumed parameter but it is extracted from the simulations. Thus,
  the derived star formation efficiency $f_*$ is a very generic
  prediction of our feedback model because it is nearly independent of
  the assumed value of $\epsilon_*$, that is instead quite uncertain.
\item {\em Contribution to Reionization}. Due to the feedback-induced
  bursting mode of star formation in pre-reionization dwarfs, the
  cosmological \HII~regions that they produce remain confined in size
  and never reach the overlap phase (\eg, Fig.~\ref{fig:pfr2}
  above). Reionization is completed by more massive galaxies.
\item {\em Gas photo-evaporation and metallicity.}  Star-forming dwarf
  galaxies show large variations in their gas content because of the
  combined effects of stellar feedback from internal sources and IGM
  reheating. Ratios of gas to dark matter lie below the cosmic mean in
  halos with masses $M_{\rm dm}<10^8$~M$_\odot$. Fig.~\ref{fig:feed}(left)
  shows the fraction of baryons retained by dark and luminous
  halos. It is clear that small mass luminous halos lose most of their
  gas before reionization due to internal radiation sources. Dark
  halos instead are able to retain gas for a longer time \citep[see
  also][]{Hoeft:06}. Gas depletion increases with decreasing redshift:
  the lower-mass halos lose all their gas first but, as the universe
  evolves, larger halos with $M_{\rm dm} \sim 10^8$ M$_\odot$ also lose a
  large fraction of their gas. Gas photoevaporation reduces the
  metallicity of pre-reionization dwarfs to levels consistent with
  observations of dSph galaxies.
\item {\em Number of luminous galaxies}. Only about 1\% of dwarf dark
  matter halos with mass $M_{\rm dm}\sim 5 \times 10^6$ M$_\odot$,
  assembled prior to reionization, are able to form stars. The fraction
  of luminous halos scales with the halo mass as $F_{\rm lum} \propto
  M_{\rm dm}^{2}$. Thus, most halos with mass $\simgt 5 \times
  10^7$~M$_\odot$ are luminous. Fig.~\ref{fig:feed}(right) shows
  $F_{lum}$ as a function of the halo mass at redshifts $z=12.5$ and
  $z=10.2$. The figure also illustrated that $f_*$ depends on the
  environment. Namely, it depends on the proximity of the
  pre-reionization dwarfs to other luminous galaxies.
\begin{figure*}[th]
\plottwo{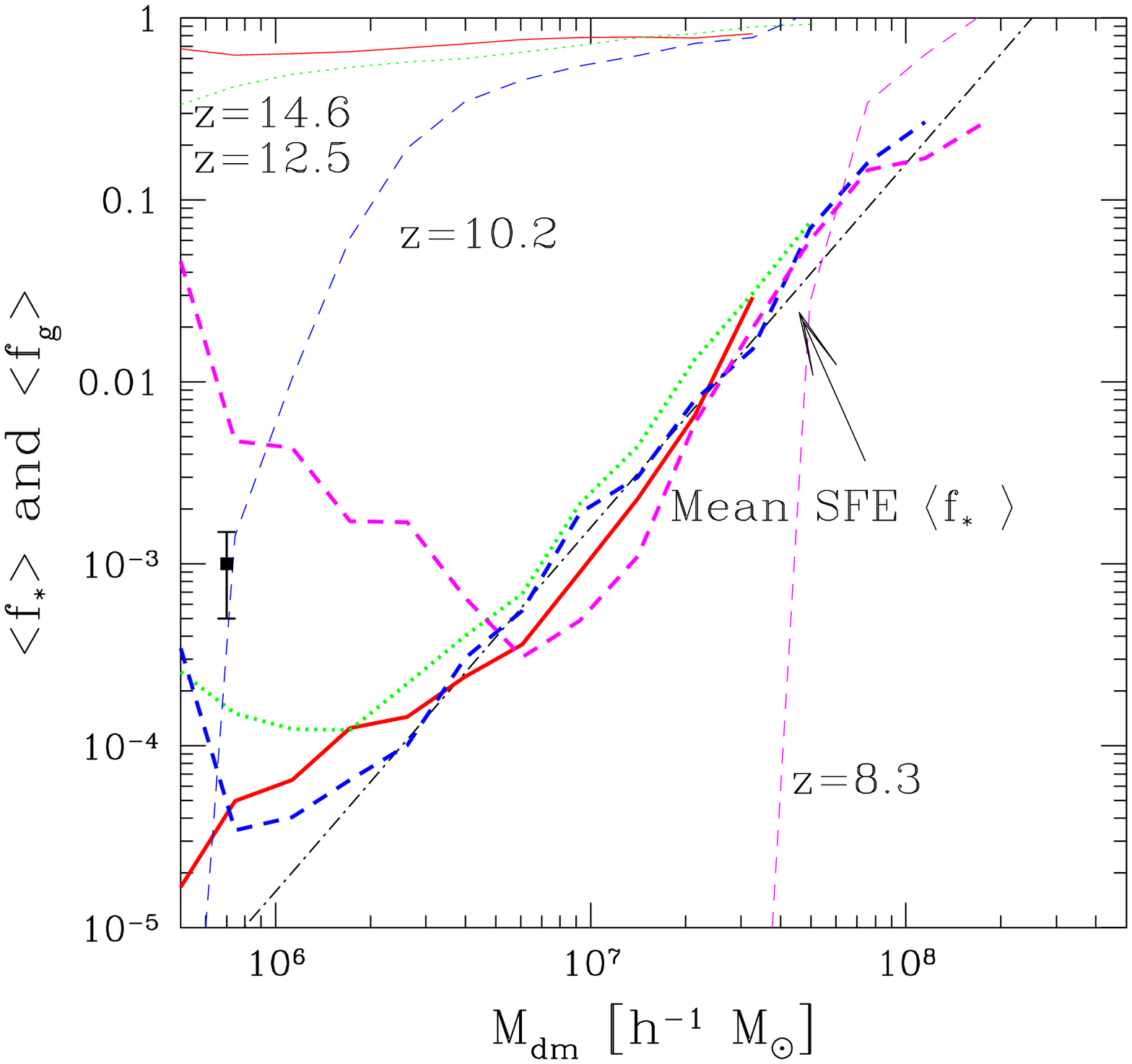}{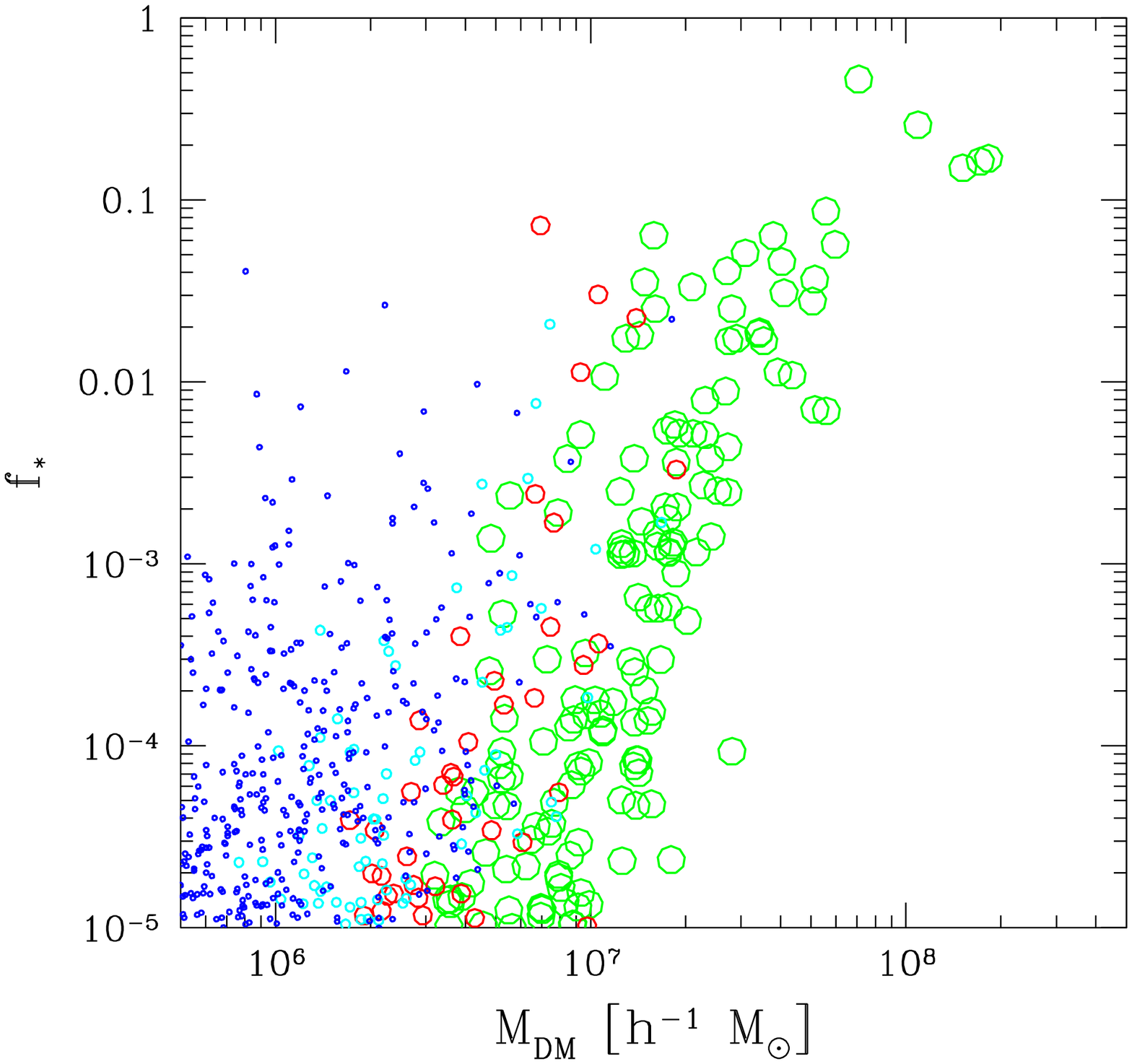}
\caption{{\it (Left)}. Average star formation efficiency (\ie,
  fraction of the collapsed baryon mass converted into stars),
  $\langle f_* \rangle=\langle M_*/M_{\rm bar}\rangle$ (thick curves),
  and gas fraction (\ie, fraction of the collapsed baryon mass
  retained in the gas phase) $\langle f_{\rm g} \rangle=\langle M_{\rm
    g}/M_{\rm bar}\rangle$ (thin curves), of the first galaxies as a
  function of their halo mass for run S2 in RGS08 at $z=14.5, 12.5,
  10.2$, and $8.3$. For comparison, the symbol with error bar shows
  the expected star formation efficiency (roughly $M_* \sim
  30-300$~M$_\odot$ divided by the baryonic mass of the halo) in the
  first mini halo of mass $10^6$ M$_\odot$ simulated by
  \cite{Abel:02}. The dot-dashed line shows a power-law fit for the
  mean SFE, $\langle f_* \rangle(M_{\rm dm},z) \propto M_{\rm
    dm}^{2}$. The SFE is nearly independent of redshift apart from an
  increase in halos with $M_{\rm dm}<10^7$~M$_\odot$ at $z\sim
  8$. {\it (Right)}. Same as the left panel but showing the star
  formation efficiency, $f_*$, for individual galaxies in the
  simulation at $z=10.2$.  Circles, from smaller to the larger, refer
  to galaxies with gas fractions $f_{\rm g}<0.1$\% (blue),
  $0.1\%<f_{\rm g}<1$\% (cyan), $1\%<f_{\rm g}<10$\% (red) and $f_{\rm
    g}>10$\% (green), respectively.}
   \label{fig:sfe}
\end{figure*}
We find $\sim 450$ dwarf galaxies per Mpc$^{3}$ with bolometric
luminosity between $10^4$ and $10^8$ L$_\odot$. The luminosity
function is rather flat at low luminosities, with about $10$ galaxies
per Mpc$^{3}$ in the range $10^7<L<10^8$ L$_\odot$, and about $220$
Mpc$^{-3}$ in the range $10^4<L<10^5$ L$_\odot$ and
$10^5<L<10^6$~L$_\odot$.
\item {\em Star formation efficiency and mass-to-light}. The mean star
  formation efficiency \hbox{$\langle f_* \rangle = \langle M_*/M_{\rm
      bar}^{\rm max} \rangle$}, where $M_{\rm bar}^{\rm max} \simeq
  M_{\rm dm}/7$, is found to be {\it nearly independent of redshift}
  and to depend on total mass as $\langle f_* \rangle \propto M_{ \rm
    dm}^{\alpha}$ with $\alpha=2$ if the radiative feedback is strong
  (\ie, top heavy IMF and/or large $\langle f_{esc}\rangle$) and
  $\alpha=1.5$ if the feedback is weak. This is shown in
  Fig.~\ref{fig:sfe}(left), where we plot the mean star formation
  efficiency, $\langle f_*\rangle $, and the mean gas fraction
  $\langle f_g\rangle$ in halos of mass $M_{\rm dm}$.
\item {\em Scatter of the mass-to-light ratio}. A tight relationship
  between the star formation efficiency $f_*$ and the total mass of
  halos holds only for galaxies with $M_{\rm dm}>5 \times 10^7$
  M$_\odot$.  In lower-mass halos, the scatter around the mean
  $\langle f_* \rangle$ is increasingly large (see
  Fig.~\ref{fig:sfe}(right)). For a given halo mass, the galaxy can be
  without stars (dark galaxy) or have star formation efficiency $f_*
  \sim 0.1$.  However, only a few dark galaxies with mass at formation
  $M_{\rm dm} \sim 1-5 \times 10^{7}$~M$_\odot$ should exist in the
  Local Group.
\item {\em Size and morphology of stellar component}. Galaxies with
  masses $M_{\rm dm}<10^8$ M$_\odot$~have a low surface brightness and
  extended stellar spheroid.  At $z \sim 10$, the outer edges of the
  stellar spheroid nearly reaches the virial radius. In more massive
  galaxies that cool more efficiently by Lyman-alpha radiation, the
  stars and light are more centrally concentrated. Fig.~\ref{fig:prof}
  shows the structural properties of the dark matter and stellar halo
  in three of the most massive galaxies in our simulation. These dwarf
  galaxies have properties similar to Draco and Umi dSphs. The figure
  also shows that the velocity dispersion of the stars in these dwarfs
  is about a factor of two smaller than $v_{\rm max}$.
\end{enumerate}
\begin{figure}[hbt]
\plotone{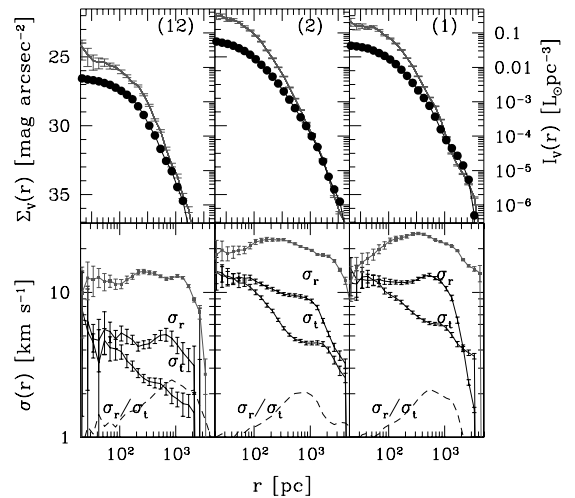}
\caption{The top panels show the surface brightness (black point) and
  luminosity density (gray points) radial profiles for 3 galaxies from
  among the most massive dwarf galaxies extracted from the RG05
  simulation at $z=8.3$. We have evolved the stellar population
  passively to $z=0$. The simulated galaxies shown in this figure have
  a stellar spheroid similar to Draco and Umi already in place at the
  time of formation (\ie, the spheroid is not produced by tidal
  effects). The bottom panels show the velocity dispersion profiles of
  the dark matter (gray points) and of the stars (black points) for
  the same 3 galaxies. The velocity dispersion of the stars is split
  in the radial and tangential components. All quantities are
  spherically averaged because the dark matter and stars have nearly
  spherical symmetry.}
   \label{fig:prof}
\end{figure}

\subsubsection{Photo-evaporation and Reionization feedback}\label{sec:reio}

The small total mass of the first galaxies has two other implications.
First, the ionizing radiation emitted by massive stars can blow out
most of the gas before SN-driven winds become important, further
reducing star formation rates (see RGS08).  Second, the increase in
temperature of the IGM to $10,000-20,000$ K due to \HI~reionization
prevents the gas from condensing into newly virialized halos with
circular velocities smaller than $10-20$~km~s$^{-1}$
\citep[\eg,][]{Babul:92,Efstathiou:92b,Bullock:00,
  Gnedin-filteringmass00, Hoeft:06, OkamotoGT08}. It follows that
dwarf galaxies with $v_{\rm max} < 10-20$~km~s$^{-1}$ lose most of
their gas before reionization and stop accreting new gas and forming
stars after reionization.

The value $v_{\rm max} \sim 20$~km~s$^{-1}$ that we use to define a
fossil is motivated by the fundamental differences in cooling and
feedback processes discussed above that regulate star formation in the
early universe. It is not the critical value for suppression of gas
accretion due to reionization.  Indeed, we discuss in
\S~\ref{sec:infall} that pre-reionization fossils may have a late
phase of gas accretion and star formation well after reionization, at
redshift $z<1-2$. Thus, a complete suppression of star formation after
reionization (about 12~Gyr ago) is not the defining property of a
fossil dwarf.

\section{Late time cold accretion from the IGM}\label{sec:infall}

\begin{figure}[t]
\plotone{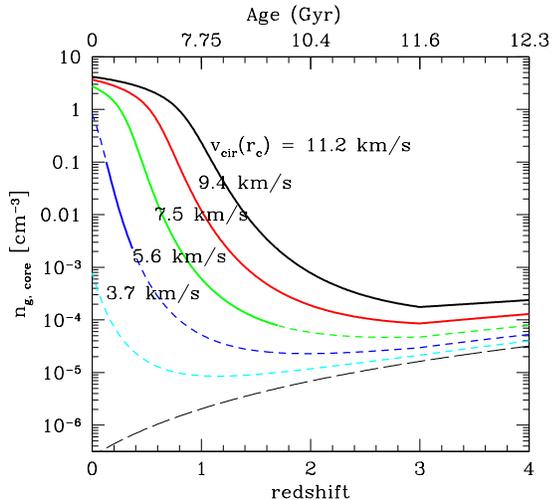}
\caption{The evolution of the gas density in the core of dark halos
  due to cold accretion from the IGM for halos with $v_{\rm vir}=18,
  15, 12, 9$ and $6$~km~s$^{-1}$ (from the top to the bottom), and
  corresponding to the circular velocities at the core radius $v_{\rm
    cir}(r_{\rm core})=0.66 v_{vir}$ (shown by the labels). The curves
  are assuming isothermal equation of state of the gas but the dashed
  portions show the parameter space in which such assumption fails
  because the gas cannot cool sufficiently fast as it is compressed
  toward the center of the halo.}\label{fig:infall}
\end{figure}

The ability of the IGM gas to condense at the center of dark halos
depends on the ratio, $\Gamma=v_{\rm vir}/c_{\rm s, igm}$, of the
circular velocity to the IGM sound speed, and also on the dark halo
concentration, $c$ \citep{Ricotti:09}. Typically, the concentration of
a halo is $c_{\rm vir} \sim 4$ at the redshift of virialization
\citep{Bullock:01, WechslerB:02} but, as the halo evolves in the
expanding universe, its concentration increases $\propto (1+z_{\rm
  vir})/(1+z)$. The evolution of the halo concentration with redshift
can be understood in the context of the theory of cosmological
secondary infall of dark matter \citep{Bertschinger:85} and has been
quantified using N-body simulations \citep{Bullock:01, WechslerB:02}.
Thus, primordial halos with $v_{\rm vir}<10-20$~km~s$^{-1}$ stop
accreting gas immediately after reionization, but, in virtue of their
increasing concentration and the decreasing temperature of the IGM at
$z<3$ (after \GII ~reionization), they may start accreting gas from
the IGM at later times \citep[see,][]{Ricotti:09}. As a result, we
expect that if the fossils of the first galaxies exist in the Local
Group (RG05), they may have a more complex star formation history than
previously envisioned. A signature of this model is a bimodal star
formation history with an old ($\sim 13$~Gyr) and a younger ($\simlt
5-10$~Gyr, depending on the halo mass) population of stars. Leo~T
properties can be reproduced by this simple model for late gas
accretion \citep{Ricotti:09}. In addition, Leo~T seems to show a
bimodal star formation history \citep{deJongetal08} as expected in our
model. Still, other models may also explain the observed star
formation history of Leo~T \citep{Stinsonetal07}.

Perhaps more important is the possible existence of dark galaxies:
small mass halos containing only gas but no stars. Dark
galaxies are most likely to exist if pre-reionization fossils do not
form efficiently due to dominant negative feedback in the early
universe (\eg, H$_2$ photo-dissociation by the FUV background).

The late gas accretion from the IGM is shown in
Figure~\ref{fig:infall} for dark halos with circular velocity at
virialization $v_{\rm vir}=18, 15, 12, 9$ and $6$~km~s$^{-1}$ (lines
from the top to the bottom). The lines show the evolution of the gas
density in the core of the halo as a function of redshift. The core
radius is typically $100$~pc and the labels show the circular velocity
at the core radius ($v_{\rm cir}(r_{\rm core}) \approx 0.66 v_{vir}
\approx 0.624 v_{\rm max}$, where $v_{\rm vir}$ is circular velocities
at the virial radius and $v_{\rm max}$ is the maximum circular
velocity) . We show the evolution of the gas density only for halos
that are affected by reionization feedback. More massive halos will
also accrete gas from the IGM as they evolve in isolation after
virialization, but the gas accretion is continuous and not affected by
reionization.
\begin{figure*}[th]
\epsscale{1.1}
%\plottwo{LeoT_f4a.eps}{LeoT_f4b.eps}
\plottwo{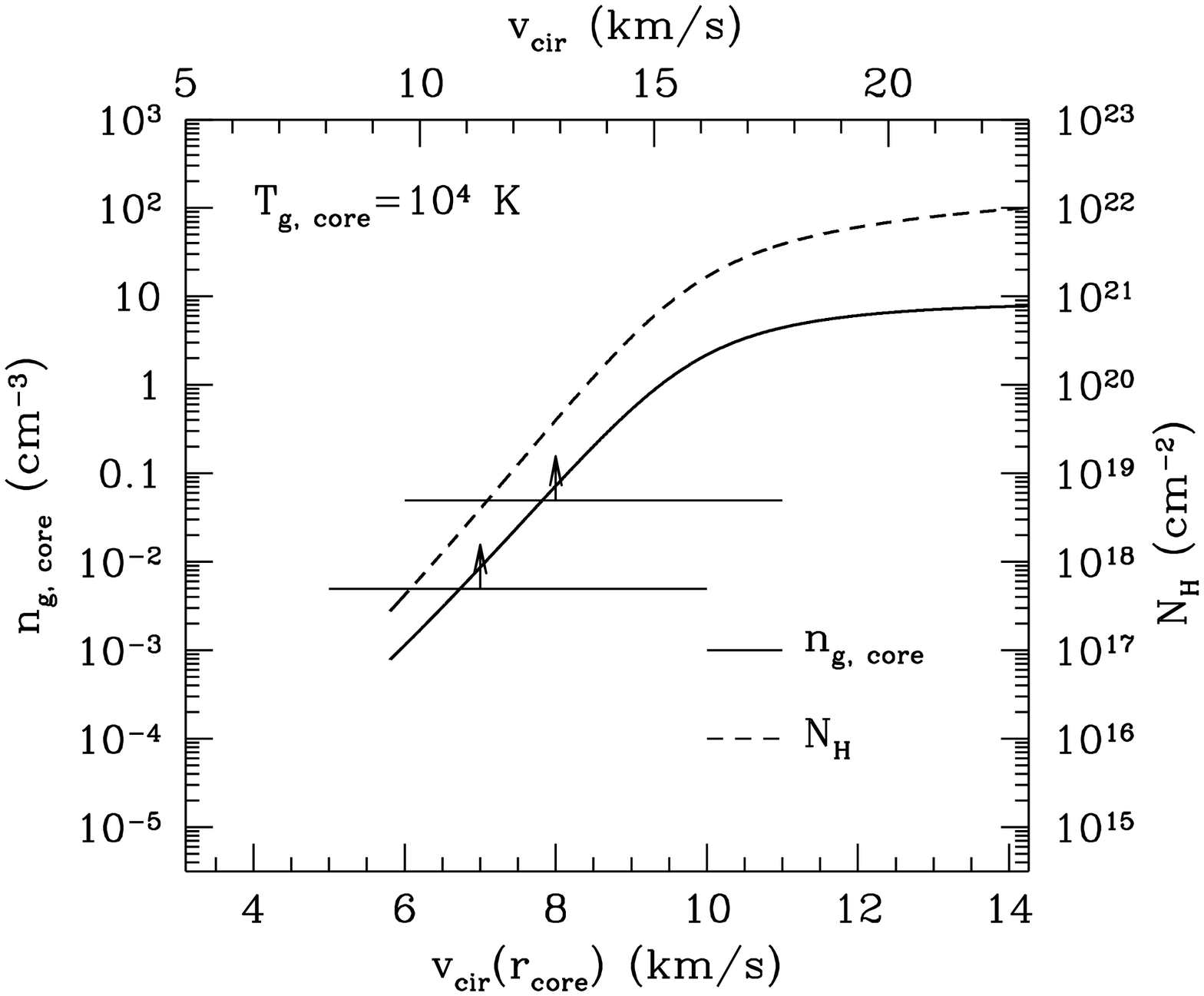}{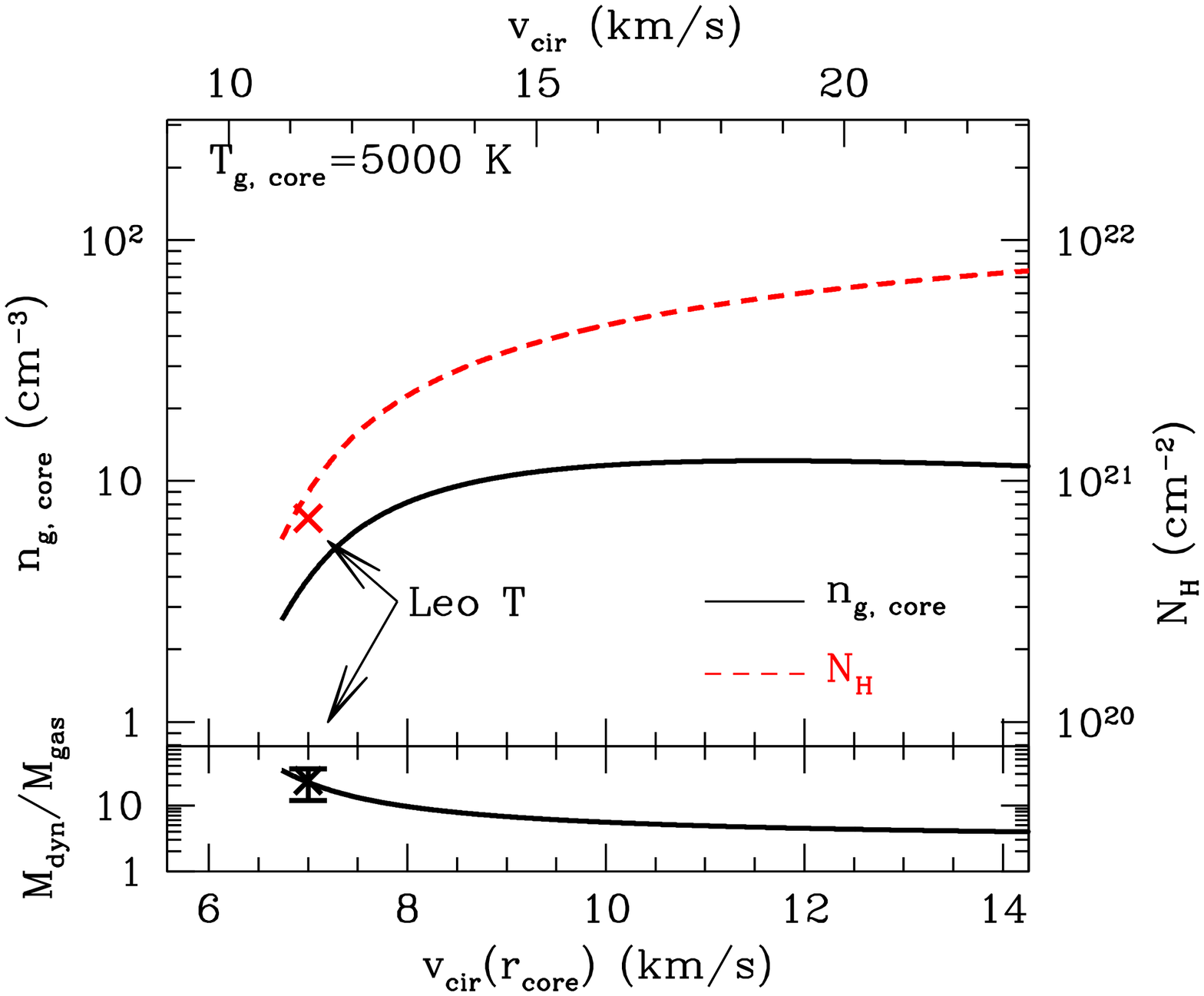}
\caption{{\it (Left).} The gas density, $n_{\rm g, core}$ (solid
  curves), and hydrogen column density, $N_{\rm H}=2r_{\rm c}n_{\rm g,
    core}$ (dashed curves), within the core, $r_{\rm c}$, of a
  minihalo at redshift $z=0$ as a function its circular velocity at
  $r_{\rm c}$. The minimum $v_{\rm cir}$ in each curve is determined
  by the condition $t_{\rm rec}/t_{\rm H}<1$, necessary for cooling to
  $T_{\rm gas} \sim 10^4$~K. The horizontal lines with arrows show the
  requirement for cooling to temperatures below $10^4$~K, necessary
  for initiate star formation, for gas metallicity $Z=0.1$ (lower
  line) and $0.01$~Z$_\odot$ (higher line). {\it (Right).}  Same as in
  the left panel but for minihalos whose gas is able to cool to
  $T_{\rm g, core}=5000$~K due to metal pre-enrichment
  ($Z=0.1$~Z$_\odot$). These halos are likely able to sustain a
  multi-phase ISM and form stars.}\label{fig:core2}
\end{figure*}

Under the scenario in which halos with masses smaller than the
critical value of $10^8-10^9$~M$_\odot$ remain dark due to feedback
effects, the increase in their dark matter concentration and the
temperature evolution of the IGM will produce a late phase of gas
accretion at redshift $z<1-2$. If the gas has very low metallicity or
is metal free, it is unlikely that the accreted gas will be able to
form stars in the smallest mass halos. This is why we envisioned the
possible existence of dark galaxies. However, their mass would be
smaller than the mass of any luminous galaxy and their discovery would
be challenging.

The level of metal pre-enrichment necessary to initiate star formation
in minihalos that experience a late phase of gas accretion can be
roughly estimated from the cooling function from hyperfine transitions
of oxygen and carbon: $\Lambda_{23} \sim 10^{-3}~(Z/Z_\odot)$, where
$\Lambda_{23}=10^{-23}$~erg~s$^{-1}$~cm$^{3}$ and $Z$ is the gas
metallicity. A necessary condition for star formation is $t_{\rm cool}
\approx (0.7~{\rm yr})~T/(n_{\rm g, core}\Lambda_{-23}) < t_{\rm H}$,
that can be written as $n_{\rm g, core} > 0.03~{\rm
  cm}^{-3}(Z/10^{-2}~Z_\odot)^{-1}$.  The left panel in
Figure~\ref{fig:core2} shows $n_{\rm g, core}$ and $N_{\rm H}$ in
minihalos that evolve isothermally at $T \sim 10^4$~K but that do not
form stars (\ie, candidates for extragalactic CHVCs and dark
galaxies). The horizontal lines show the requirement for metal
cooling and star formation assuming gas metallicity $Z=0.1$ and
$0.01$~Z$_\odot$. The right panel in Figure~\ref{fig:core2} shows
$n_{\rm g, core}$, $N_{\rm H}$ and $M_{\rm dyn}/M_{\rm gas}$ (the
dynamical mass to gas mass ratio) in the core of minihalos that are
able to cool to $T_{\rm g, core}=5000$~K (roughly the temperature of
the ISM in Leo~T), and thus form stars. The symbols show the observed
value for Leo~T.

\section{Comparison of theory and observations}\label{sec:comp}

\subsection{Number of fossils and non-fossil satellites in the Milky
  Way}\label{ssec:count}

N-body simulations can be used to estimate the number of dark halos in
the Milky Way with maximum circular velocity $v_{\rm
  max}>20$~km~s$^{-1}$.  However, there is a complication to this
naive calculation. A significant fraction of dark halos that today
have $v_{\rm max} < 20$~km~s$^{-1}$ were once more massive, due to
tidal stripping \citep{KravtsovGnedinKlypin04}.  According to our
definition, dwarf galaxies formed in these dark halos would not be
pre-reionization fossils if they had at any time during their
evolution $v_{\rm max}(t)>20$~km~s$^{-1}$ (see \S~\ref{sec:reio}).  If
the number of observed Milky Way satellites exceeds the estimated
number of these massive halos we must conclude that at least a
fraction of the observed Milky Way satellites are pre-reionization
fossils.

However, there is an assumption in this scenario. One must assume that
the stars in these halos survive tidal stripping for as long as the
dark matter. In this case tidally stripped halos may indeed account
for a fraction or all of the newly discovered ultra-faint dwarfs.
However, \cite{PenarrubiaNM08} find that tidally stripped dark halos
lose their stars more rapidly than they lose their dark matter. Thus,
they may become dark halos even though they were initially luminous
satellites. These dark halos should not be counted as ultra-faint
dwarfs.

Using results of published N-body simulations of the Milky Way,
\cite{BovillR:09} have estimated the number of dark halos that have or
had in the past $v_{\rm max}(t)>20$~km~s$^{-1}$ (\ie, non
pre-reionization fossils).  In Table~\ref{tab:count} we summarize the
results of the counts for dark matter and luminous satellites for two
large N-body simulations of a Milky Way type halo: the ``Aquarius'
simulation \citep{Springeletal08} and the Via Lactea I simulation
\citep{Diemandetal07b}.
\begin{table}[tbh] 
\caption{Number of observed satellites versus number of dark halos with $v_{\rm max}(t)>20$~km~s$^{-1}$ (\ie, non pre-reionization fossils) for the Milky Way}
\centering 
\begin{tabular}{lccccc}
\hline\hline                        %inserts double horizontal lines 
Distance & Luminous & \multicolumn{4}{c}{Dark halos with $v_{\rm max}(t)>20$~km~s$^{-1}$}\\ 
from & dwarfs& 
\multicolumn{2}{c}{Via Lactea I sim.} &
\multicolumn{2}{c}{Aquarius sim.}\\
center& & today & any time & today & any time\\[0.5ex]
\hline                    % inserts single horizontal line 
$<200$~kpc & 176 to 330 & 14 & $36 \pm 8$ & 34 & $91 \pm 20$ \\
$<417$~kpc & 304 to 576 & 28 & $73 \pm 16$ & 69 & $182 \pm 40$ \\[1ex]
\hline     %inserts single line 
\end{tabular} \label{tab:count}  % is used to refer this table in the text 
\end{table} 

The number of luminous satellites that exist within the Milky Way is
highly uncertain beyond a distance from the Galactic center of
$200$~kpc. \cite{Tollerudetal08}, after applying incompleteness
corrections, estimated \hbox{304--576} satellites within $417$~kpc and
about \hbox{176--330} within $200$~kpc (the numbers are from their
Table~3).  As shown in Table~\ref{tab:count}, the existence of some
pre-reionization fossils among the ultra-faint dwarfs appears to be
favored by the data.  However, the current uncertainties on the
completeness corrections of observations and on the simulations are
too large to deem the existence of fossils as necessary.

The error bars on the theoretical estimate of the number of fossils in
the Milky Way shown in Table~\ref{tab:count} come from uncertainties
in the fraction of halos that were more massive in the past. This
fraction was derived from simulations by
\citep{KravtsovGnedinKlypin04}. Another uncertainty in the simulation
results can be attributed to the different predictions for the number
of Milky Way satellites in the Via Lactea I and II and Aquarius
simulations. The discrepancy can be partially attributed to different
cosmology in the simulations but mostly because the Via Lactea I
simulation likely used erroneous initial conditions.  Finally,
\cite{Tollerudetal08} corrections on the number of observed satellites
also rely on the radial distribution of dark matter sub halos
extracted from Via Lactea I simulations that may be erroneous. Once
the discrepancies among different simulations are better understood
the number of simulated satellites of the Milky Way may be known with
greater certainty.

Using comparisons between the predicted and observed Galactocentric
distributions of dwarf satellites around the Milky Way,
\cite{GnedinKravtsov06}, hereafter GK06, have estimated that
pre-reionization fossils may constitute about $1/3$ of Milky Way
dwarfs. GK06 estimated the number of fossils in the Milky Way using
data from the simulations of the first galaxies in RG05. GK06 defined
a fossil as a simulated halo which survives at $z = 0$ and remains
below the critical circular velocity of $20$~km~s$^{-1}$ with no
appreciable tidal stripping (the usual definition of fossil adopted in
this paper as well). They calculate the probability, $P_S(v_{\rm max},
r)$, of a luminous halo with a given maximum circular velocity $v_{\rm
  max}$ to survive from $z = 8$ (the final redshift of the RG05
simulation) to $z = 0$. The surviving halos are assigned a luminosity
based on the $L_{\rm V}$ versus $v_{\rm max}$ relationship from RG05.
At $z = 0$, GK06 has a population of dwarf galaxies with a resolution
limit of $v_{\rm max} = 13$~km~s$^{-1}$.  This limit corresponds to a
lower luminosity limit of $L_{\rm V} \sim 10^5$~L$_{\odot}$, which
includes Leo T and Canes Venatici I, but excludes all the other new
ultra-faint Milky Way satellites.

\begin{figure}[t]
\plotone{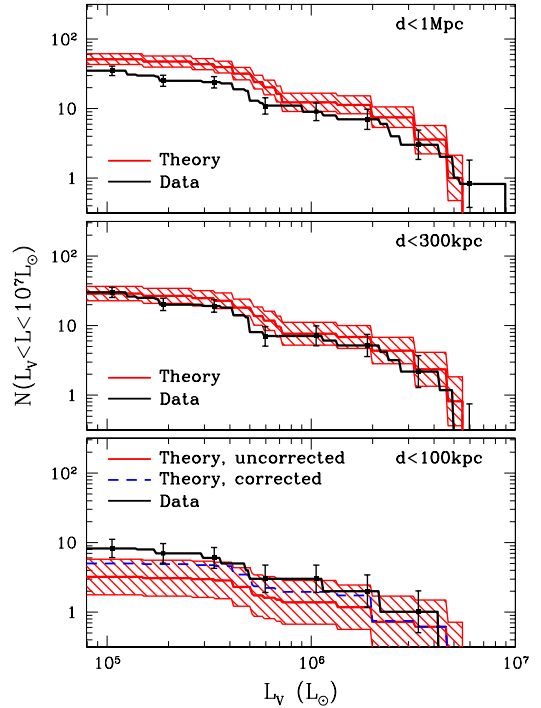}
\caption{Luminosity function of pre-reionization fossil dwarfs
  predicted in GK06 (red bands) plotted with the luminosity function
  for Local Group dSphs (points with error bars).  The data from
  observations are corrected only for limits in sky coverage of the
  SDSS survey.}
   \label{LF}
\end{figure}

In Figure~\ref{LF}, we show the cumulative luminosity function from
GK06 for the Milky Way and M31 satellites.  The lower panel shows
satellites with distance from their host $d<100$~kpc, the middle panel
$d<300$~kpc and the upper panel $d<1$~Mpc.  The gray lines show the
GK06 predictions, and the shaded region encompasses the error bars.
The resolution limits in GK06 cause halos with $v_{ \rm
  max}<17$~km~s$^{-1}$ to be preferentially destroyed by tidal
effects. The dashed line show the predicted luminosity function
corrected for the resolution effects.  Both the uncorrected (solid
lines) and corrected (dashed lines) luminosity functions are plotted
in the lower panel. The points with error bars show the observed
luminosity function of dSph galaxies around the Milky Way and M31
corrected only for limits in sky coverage of the SDSS survey. The plot
is from GK06 but has been updated to include the new ultra-faint
dwarfs with $L_{\rm V} \simgt 10^5$~L$_\odot$.

The results of this model are consistent with the observations. The
model reproduces the Galactocentric distribution of the most luminous
dSphs, even though in this model dSphs are not tidally stripped
dIrrs. It also shows a good agreement with observations for
luminosities that can be considered nearly complete within a given
Galactocentric distance.

\subsection{Statistical Properties of  pre-reionization Fossils}\label{sec:prop}

\begin{figure*}[th]
\plottwo{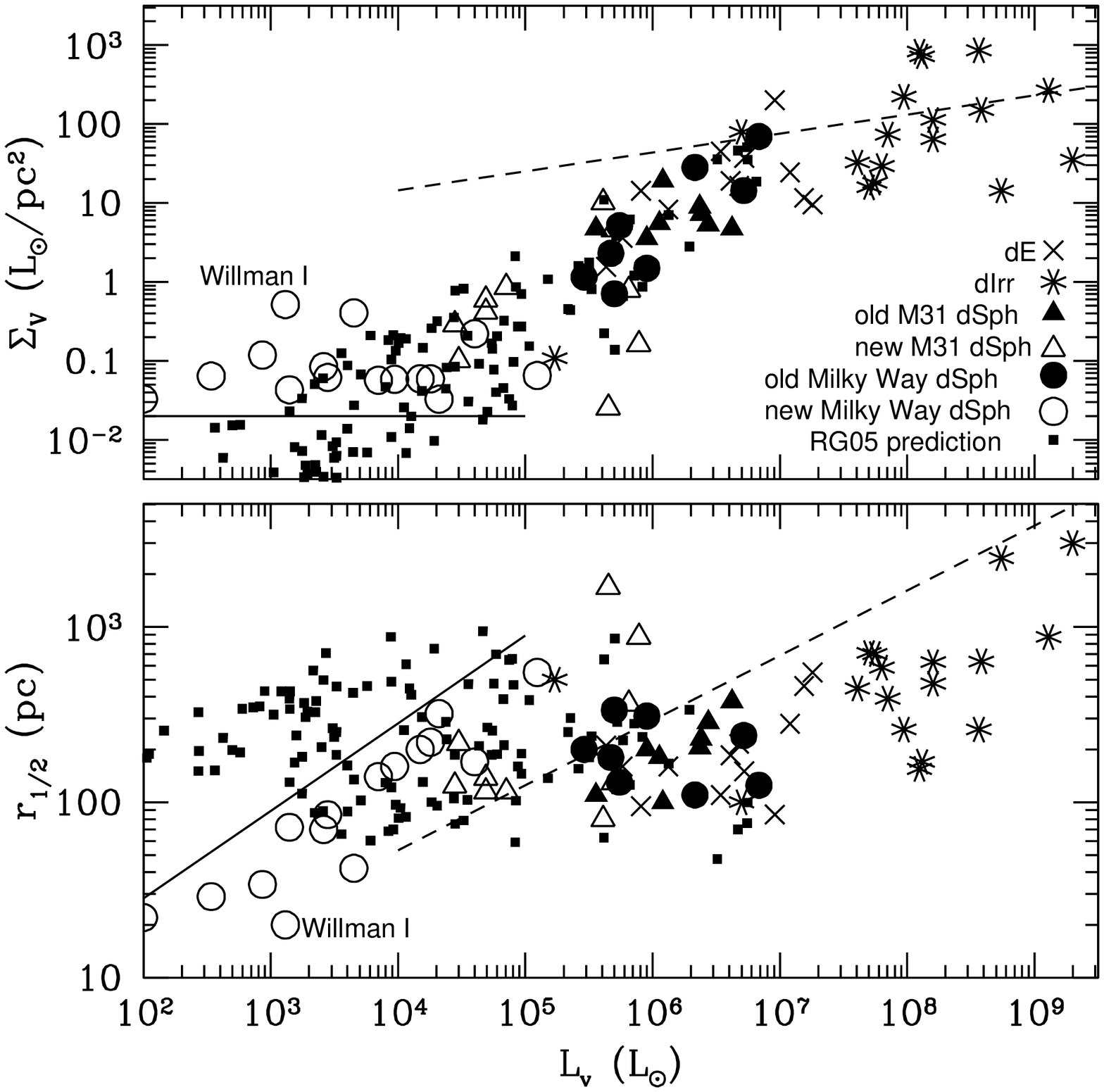}{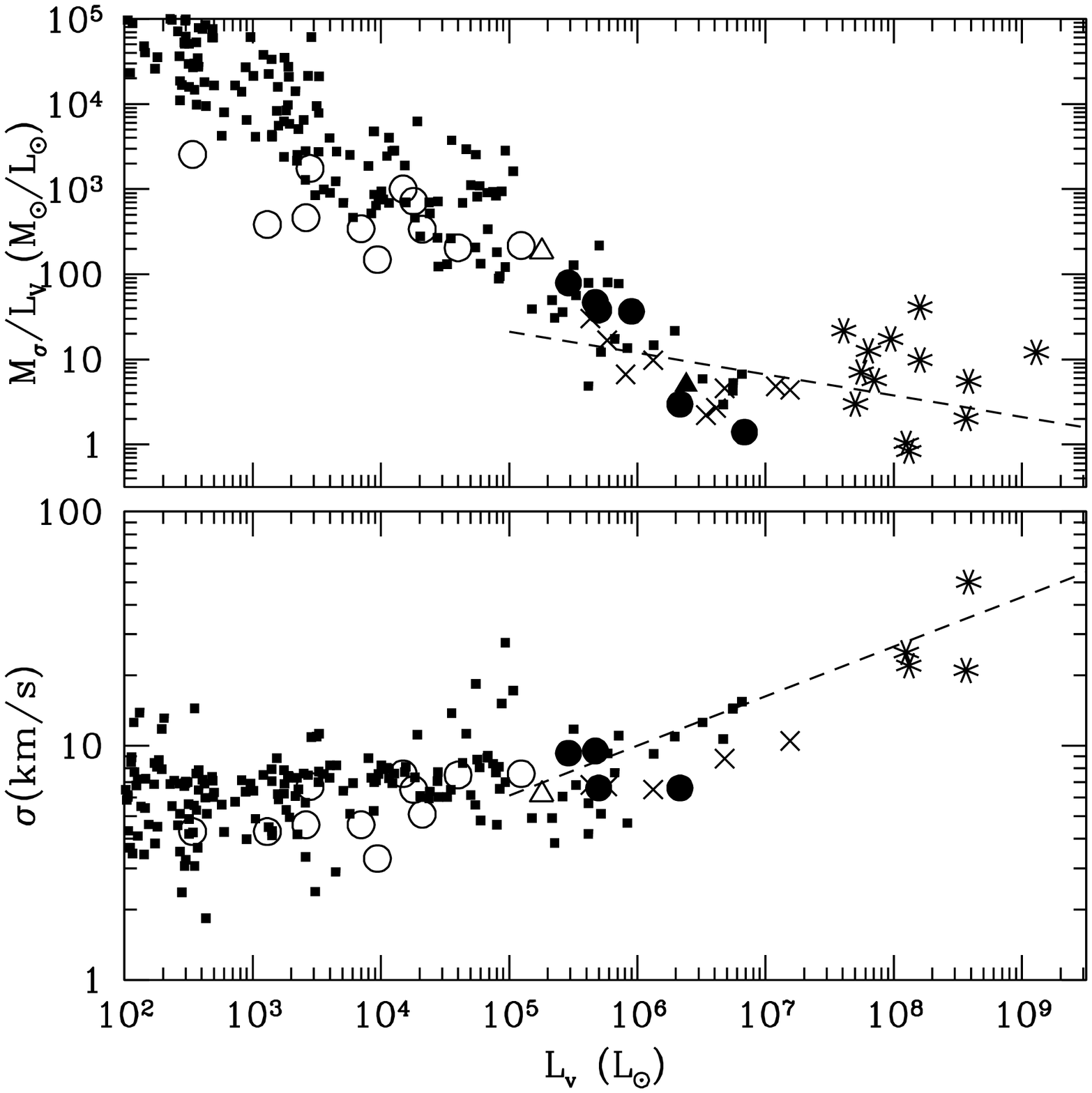}
\caption{{\it (Left).} Surface brightness and core radius vs. V-band
  luminosities.  Small filled squares are simulated pre-reionization
  fossils from RG05, asterisks are dIrrs, crosses are dEs, closed
  circles are the previously known dSph around the Milky Way, closed
  triangles are previously known dSph around M31, and open circles and
  triangles are new dSph around the Milky Way and M31
  respectively. The solid lines roughly show the detection limits
  inherent to the methods used to find the ultra-faints in the SDSS
  data \citep{Koposov:08} and the dashed lines show the scaling
  relationships for more luminous Sc-Im galaxies ($10^8L_\odot \simlt
  L_{\rm B} \simlt 10^{11}L_\odot$) derived by \cite{Kormendy:04}.{\it
    (Right).}  Mass-to-light ratio and velocity dispersion of a subset
  of the new dwarfs \citep{Martinetal07,SimonGeha07} vs V-band
  luminosity. The symbols and lines are as in the left panel.}
   \label{Kor}
\end{figure*}

In this section, we compare the properties of the new dwarf galaxies
discovered in the Local Group to the theoretical predictions of
simulations of primordial galaxies formed before reionization. The
argument that justifies this comparison is that star formation stops
or is greatly reduced after reionization (but see
\S~\ref{sec:infall}).  We do not expect two perfectly distinct
populations of fossil galaxies with $v_{\rm max}<20$~km~s$^{-1}$ and
non-fossils with $v_{\rm max}\ge 20$~km~s$^{-1}$, but a gradual
transition of properties from one population to the other.  Some
fossils may become more massive than $v_{\rm max} \sim 20$~km~s$^{-1}$
after reionization, accrete some gas from the IGM, and form a younger
stellar population. These dwarfs are no longer defined as
``fossils''. However, if the dark halo circular velocity remains close
to $20$~km~s$^{-1}$ the young stellar population is likely to be small
with respect to the old one. In RG05 we call these galaxies ``polluted
fossils'' because they have the same basic properties of ``fossils''
with a sub-dominant young stellar population.  A similar argument can
be made regarding the late phase of gas accretion that may produce
objects similar to Leo~T.

\begin{figure*}[th]
%   \resizebox{6in}{!}{\includegraphics{ZL.update.ps}}
\plottwo{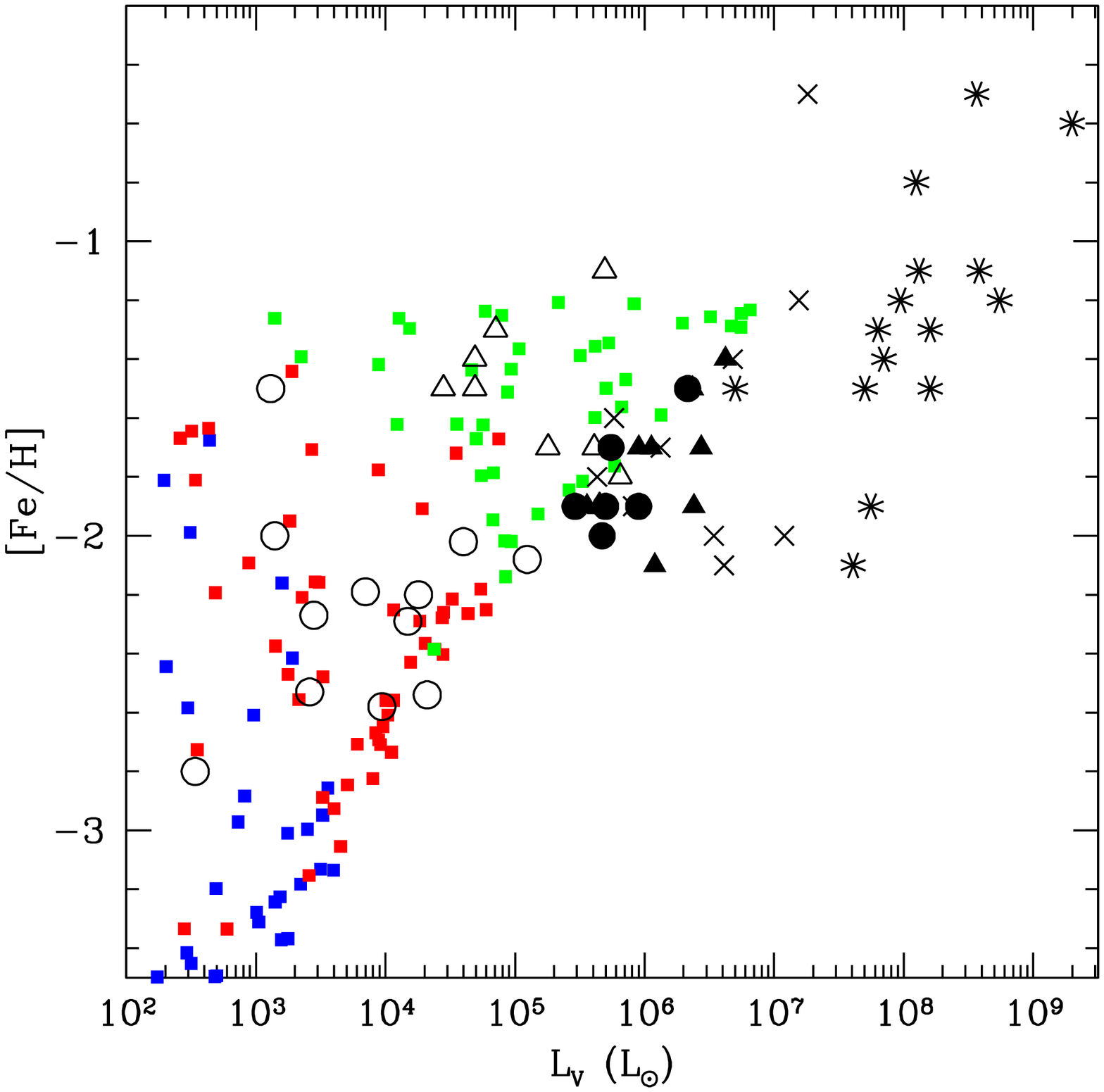}{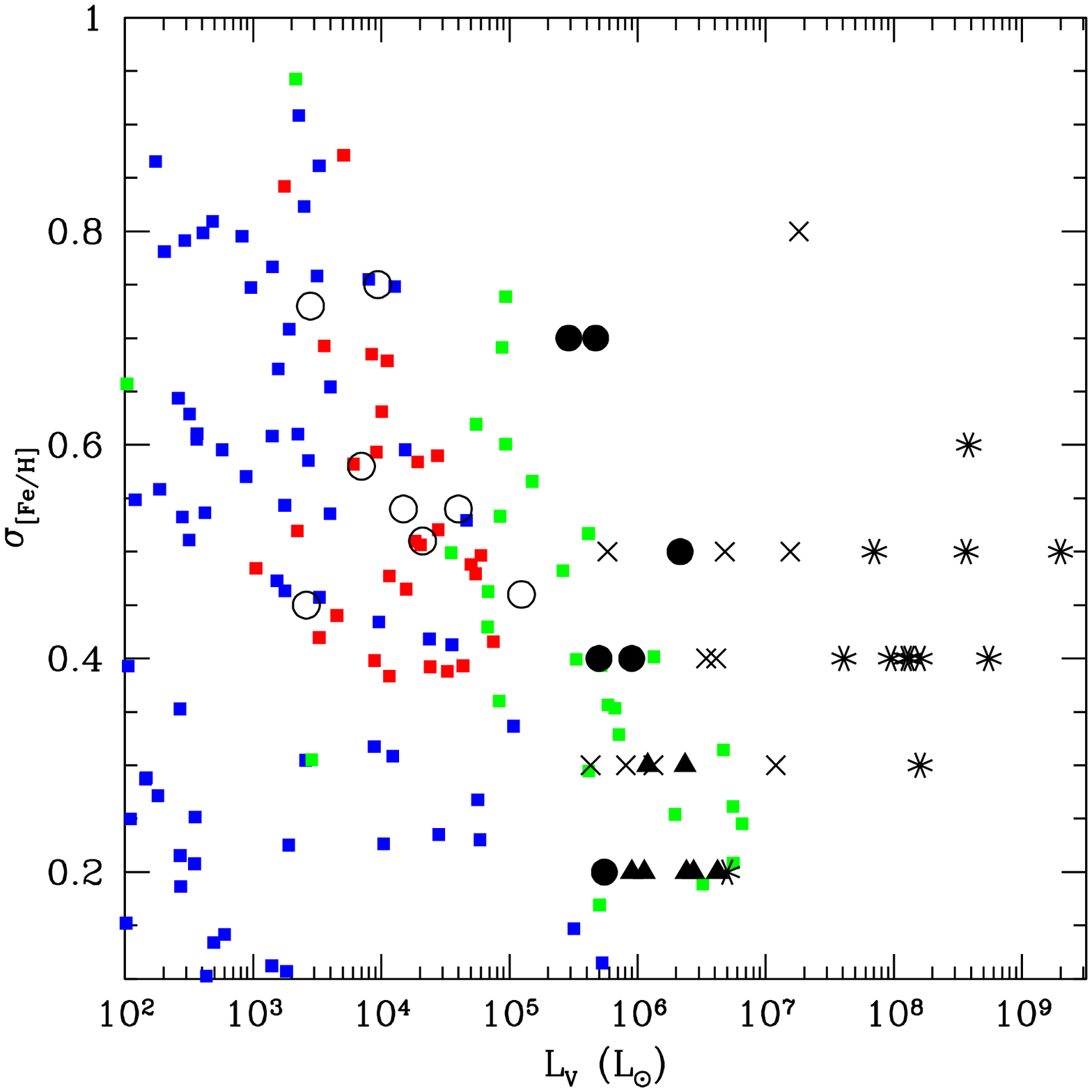}
\caption{{\it (Left).} Mean metallicity of the stars vs. V-band
  luminosity for Local Group dSphs plotted against RG05
  predictions. They symbols have the same meaning as in Fig.~8. In
  this plot simulated fossils (shown as small squares) are color coded
  according to their star formation efficiency, $f_*$: red symbols
  show simulated dwarfs with $f_*<0.003$, blue $0.003 \le f_* \le
  0.03$ and green $f_*>0.03$. {\it (Right).} Same as the left panel
  but showing the spread of the stellar metallicities in each dwarf
  (\ie, variance of the metallicity distribution) vs. their V-band
  luminosity.}
   \label{ZL}
\end{figure*}

In Figs.~\ref{Kor}-\ref{ZL} we compare the RG05 predictions for the
fossils of primordial galaxies to the observed properties of the new
Milky Way and M31 dwarfs. The symbols and lines in
Figs.~\ref{Kor}-\ref{ZL} have the following meanings.  All known Milky
Way dSphs are shown by circles; Andromeda's dSphs satellites are shown
by triangles; simulated fossils are shown by the small solid squares.
The solid and open symbols refer to previously known and new dSphs,
respectively. The transition between fossils and non-fossil galaxies
is gradual. In order to illustrate the different statistical trends of
``non-fossil'' galaxies we show dwarf irregulars (dIrr) as asterisks
and the dwarf ellipticals (dEs) as crosses, and we show the
statistical trends for more luminous galaxies as thick dashed lines on
the right side of each panel.

Figure~\ref{Kor}(left) shows how the surface brightness (top panel)
and half light radius (bottom panel) of all known Milky Way and
Andromeda satellites as a function of V-band luminosity compares to
the simulated fossils.  The surface brightness limit of the SDSS is
shown by the thin solid lines in both panels of the figure.  The new
dwarfs agree with the predictions up to this threshold, suggesting the
possible existence of an undetected population of dwarfs with
$\Sigma_{\rm V}$ below the SDSS sensitivity limit.  The new M31 satellites
have properties similar to their previously known Milky Way
counterparts (\eg, Ursa Minor and Draco).  Given the similar host
masses and environments, further assuming similar formation histories
for the halos of M31 and the Milky Way, we may be tempted to speculate
on the existence of an undiscovered population of dwarfs orbiting M31
equivalent to the new SDSS dwarfs.

The large mass outflows due to photo-heating by massive stars and the
subsequent suppression of star formation after an initial burst make
reionization fossils among the most dark matter dominated objects in
the universe, with predicted mass-to-light ratios as high as $10^4$
and $L_{\rm V} \sim 10^3-10^4$~L$_{\odot}$.  Figure~\ref{Kor}(right)
shows the velocity dispersion (bottom panel) and mass-to-light ratios,
$M_\sigma/L_{\rm V}$ (top panel), as a function of V-band luminosity
of the new and old dwarfs from observations in comparison to simulated
fossils.  The symbols are the same as in the previous
figures. Theoretical and observed dynamical masses are calculated from
the velocity dispersions of stars (\ie,
$M_\sigma=2r_{1/2}\sigma^2/G$), and do not necessarily reflect the
total mass of the dark halo at virialization.

Observations show that the value of the dynamical mass within the
stellar spheroid, $M \sim (1 \pm 5) \times 10^7$~M$_\odot$, remains
relatively constant as a function of $L_{\rm V}$
\citep{Mateo98}. Recent work by \cite{Strigarietal08} shows analogous
results to the one found by \cite{Mateo98}. The dynamical mass of
dwarf spheroidals within a radius of 300~pc is relatively constant: $M
\sim 10^7$~M$_\odot$. The radii of the stellar spheroids in these
dwarf galaxies may be either larger or smaller than 300~pc. In the
later case, the determination of the mass of the dwarfs is uncertain.

Our simulation provides some insight into the reason why the dynamical
mass remains roughly constant in dSphs.  The simulations show that in
pre-reionization dwarfs, the ratio of the radius of the stellar
spheroid to the virial radius of the dark halo decreases with
increasing dark halo mass (\ie, the stellar profile becomes more
concentrated for more luminous dwarfs).  Thus, as the halo mass and
virial radius increases, the stellar spheroid becomes increasingly
concentrated in the deepest part of the potential well. If follows
that the ratio, $f_\sigma \equiv M_\sigma/M_{\rm dm}$, of the
dynamical mass within the largest stellar orbits to total dark matter
mass is also reduced. Thus, the decrease of $f_\sigma$ for increasing
dark matter mass of halos maintains the value of the dynamical mass
$M_\sigma = f_\sigma M_{\rm dm}$ (measured by the velocity dispersion
of the stars) almost constant, even though the total mass of the halo
increases. The extent of the stellar spheroids in the lowest mass
dwarfs is comparable in size to their virial radii at formation (see
\S~\ref{sec:sims}).

The metallicity-luminosity relation of the observed and simulated
dwarfs is shown in the left panel of Figure~\ref{ZL}.  [Fe/H] is
plotted against V-band luminosity in solar units.  Symbols for the
previously known dwarfs, the new, ultra-faint dwarfs, and simulated
fossils are the same as in Figure~\ref{Kor}.  In this plot we color
code simulated fossils according to their star formation efficiency,
$f_*$. Red symbols show simulated dwarfs with $f_*<0.003$, blue $0.003
\le f_* \le 0.03$ and green $f_*>0.03$.

Using the data for the metallicity collected in \cite{BovillR:09}, the
new ultra-faint dwarfs do not appear to follow the tight
luminosity-metallicity relationship observed in more luminous galaxies
(although error bars are large). Note that here, as well as in
\cite{BovillR:09} (although Table~3 in that paper was erroneously not
updated), we have plotted data from \cite{Kirby:08} for the 6
ultra-faint Milky Way satellites provided in that paper.  There are
several physical mechanisms that may produce the observed scatter in
metallicities of dwarfs at a given constant luminosity.  The large
spread of star formation efficiencies producing a dwarf of a given
luminosity in our simulations is responsible for at least part of the
large spread of the luminosity-metallicity relation. At this point it
is unclear whether our simulations can reproduce the scatter of
metallicities of simulated fossils, or if perhaps the luminosity of
the lowest luminosity ultra-faint dwarfs has been reduced due to tidal
interactions. As mentioned before we have suggested that the lowest
luminosity ultra-faint dwarfs have not yet been discovered because
their surface brightness lies below the SDSS detection limits.

Figure~\ref{ZL}(right) shows the scatter of the metallicity of the
stars, $\sigma_{\rm [Fe/H]}$, plotted against V-band luminosity and
[Fe/H] respectively.  The various point types and colors are the same
used in the left panel of figure~\ref{ZL}. The large spread in the
metallicity of the stars is a natural consequence of the hierarchical
assembly of dwarf galaxies in cosmological simulations. It is not
necessarily an indication that star formation in dwarf satellites was
protracted for longer than $1$~Gyr, as argued in \cite{Grebel:04} to
prove that star formation in dwarf spheroidals is not stopped by
reionization feedback.

\section{Discussion: The tidal scenario vs fossil hypothesis}\label{sec:disc}

According to the results summarized above in Table~\ref{tab:count},
the number of dark matter satellites of the Milky Way that have or had
in the past $v_{\rm max}>20$~km~s$^{-1}$ is smaller than the number of
observed luminous satellites (after applying completeness
corrections). This implies that non-fossil galaxies cannot account for
all the observed satellites. Thus, pre-reionization fossils should
exist.

However, we have already discussed the several uncertainties in
estimating the numbers summarized in Table~\ref{tab:count}. Additional
uncertainties that render the identification of fossils uncertain are
the following. The mass and circular velocity of the dark matter halo
of the Milky Way are not known precisely. Simulations should take into
account the effect of baryons in modifying the density profile and the
radial distribution of satellites. The effect of tidal stripping on
the properties of the stars in the satellites is not well understood,
thus we do not know if the tidal scenario is consistent with
observations of ultra-faint dwarfs. The luminosity and stellar
properties of non-fossil dwarf satellites are not known.

Non-fossil galaxies with $v_{\rm max}>20$~km~s$^{-1}$ may lose a
substantial fraction of their mass due to tidal interactions.  If they
survive the interaction, their properties, such as surface brightness
and half light radius, may be modified. \cite{KravtsovGnedinKlypin04}
estimate that about 10$\%$ of Milky Way dark matter satellites were at
least ten times more massive at their formation than they are
today. Although their simulation does not include stars, they favor
the idea that the stellar properties of these halos would remain
unchanged (\ie, dSphs are not tidally stripped dIrr). In their model
the majority of brighter dwarf satellites have been considerably more
massive in the past and could have formed their stars undisturbed
after reionization. More precisely, the redshift of reionization does
not affect the results of their model for classic dwarfs because the
probability of these to be fossils is low.

This version of the tidal model may be hard to distinguish
observationally from the model we propose for the fossils because in
both models the properties of the dwarfs are not modified by tidal
forces (\ie, their properties are those at formation). In addition,
fossil dwarfs may stop forming stars for only about 2 to 4 Gyr after
reionization, before starting to accrete gas again from the IGM. Thus,
reionization may imprint a bimodal star formation history in some
fossils, but this signature is not a robust discriminant because the
star formation history of dwarfs cannot be determined with sufficient
accuracy.

Observations seem to suggest that star formation in dwarf galaxies
slightly more massive than $10^8-10^9$~M$_\odot$ may be similar to
star formation in fossils and thus fit the observed properties of
classical dSphs without requiring significant tidal stripping of
stars. If star formation was included in
\cite{KravtsovGnedinKlypin04}, their model may have reproduced the
properties of observed dwarf satellites that our simulations of
pre-reionization dwarf galaxies already does. The differences between
the two models will depend on whether fossil galaxies are allowed to
form and on their properties.  In other words, the two models may
differ on the assumed mass of the smallest dark halo that can host
luminous satellites. This critical mass cannot be directly observed in
dwarf galaxies but, in principle, can be constrained by comparing the
observed number of luminous satellites to the model
predictions. Determining the minimum mass for a dark halo to become
luminous is of great importance in understanding galaxy formation in
the early universe.

To summarize, there are a few observational tests that can be used to
distinguish true fossils from dSphs or dEs that form in more massive
halos and form stars unaffected by IGM reionization. True fossils
should have either a single old stellar population or have a bimodal
star formation history produced by a temporary suppression of star
formation after reionization and late gas accretion.  In addition, if
the number of observed Milky Way satellites (or the number of isolated
dwarfs) exceeds some critical value determined using N-body
simulations (\eg, see Table~\ref{tab:count}), we may conclude that
some pre-reionization fossils do exist in the Local Group.

It is likely that these tests will prove inconclusive for some time to
come, unless the number (after corrections for completeness) of new
ultra-faint galaxies surges in the coming years. The weakness of the
star formation history test is that it requires measurements with
precision of 1-2~Gyr of the stellar populations in order to be really
discriminating between models that are quite similar to each
other. This is hard to achieve especially for ultra-faint dwarfs with
few stars.  If the number of ultra-faint dwarfs remains about the same
as today, the number argument may also remain controversial until more
detailed theoretical modeling can reduce the current uncertainties
surrounding the expected number of dark halos in the Milky Way and the
completeness corrections of the observations.  Ultimately, the case
for the origin of ultra-faint dwarfs must be made on the basis of the
model that does the best job of reproducing available observations.

Finally, even if pre-reionization fossils do not exist (\ie, halos
with $v_{\rm max}<20$~km~s$^{-1}$ are all dark), a fraction of them
should be able to accrete some gas at redshift $z<1-2$ and might be
discoverable in the outer parts of the Local Group using H$\alpha$ or
21~cm surveys \citep[\eg, ALFALFA survey,][]{Giovanelli:05,
  Giovanelli:07}. Of course, one should prove that the gas clouds are
embedded in dark halos. Measurements of the gas cloud size, column
density, and velocity broadening of the emission/absorption lines can
be used to discriminate between ``dark galaxies'' and tidal
debris. This is because the gas in dark galaxies is confined by the
gravitational potential of the dark matter halo, while tidal debris or
clouds formed via thermal instability are confined by the external gas
pressure \citep{Ricotti:09}. This is another promising direction for
determining the minimum mass of luminous galaxies in the universe.

Another variant of the tidal hypothesis for the origin of dSphs is a
scenario in which dIrr galaxies transform into dSphs as they fall into
the Milky Way and Andromeda, due to tidal and ram pressure stripping
\citep[][]{Mayeretal07,Mayeretal06}.  A work by \cite{PenarrubiaNM08}
explores the idea that ultra-faint dSphs are tidally stripped
dIrrs. They achieve some success in reproducing observed properties of
ultra-faint dwarfs. While this type of tidal stripping can reproduce
properties of an individual galaxy, it is unable to completely
reproduce all the trends in the ultra-faint population.  This is
primarily seen in the kinematics of the ultra-faint dwarfs.  Tidal
stripping predicts a steeper than observed drop in the velocity
dispersion of the stars with decreasing $L_V$ \citep{PenarrubiaNM08}.
In addition several dSph do not show signs of strong tidal stripping.
And XII and And XIV may be on their first approach to the Local Group
\citep{Martinetal06, Chapmanetal07}. Other examples of dSphs that are
found distant from the center of their host galaxies are And XVIII,
Cetus and Tucana \citep{McConnachieetal08}.

Finally, another interesting case study is Leo~T, that we have
discussed extensively above in \S~\ref{sec:leoT} and
\S~\ref{sec:infall}. Leo~T properties can be explained in some detail
as being a fossil that experienced a late phase of gas accretion
\citep{Ricotti:09}. However, another possibility that should be
explored quantitatively with simulations is that Leo~T is more
massive than a fossil but less massive than dIrr galaxies.

\section{Conclusions and Future work}\label{sec:conc}

We have summarized our work on the formation of the first galaxies
before reionization (\ie, pre-reionization dwarfs) and the quest to
identify the fossils of these first galaxies in the Local Group.  The
definition of a pre-reionization fossil is not directly related to the
suppression of star formation experienced by these galaxies due to
reionization feedback. Indeed, we discussed how pre-reionization
fossils may experience a late phase of gas accretion and possibly star
formation at redshift $z<1-2$. Most importantly, fossils are a
population of dwarf galaxies whose formation (\ie, the fraction of
halos that are luminous) is self-regulated on cosmological distance
scales by radiative processes. Their existence is not certain due to a
possible strong negative feedback that may prevent the majority of
these halos from ever forming stars. In addition, if negative feedback
heavily suppresses the number and luminosity of these first galaxies,
more massive halos with $v_{\rm max}>20$~km~s$^{-1}$ will evolve
differently because of the lower level of metal pre-enrichment of the
IGM.  To summarize, the critical circular velocity $v_{\rm max} \sim
20$~km~s$^{-1}$ that we adopt to define a fossil is primarily
motivated by fundamental differences in cooling and feedback processes
that regulate star formation in these halos in the early
universe. However, it is also close to the critical value for
continued gas accretion after IGM reionization
\citep{Gnedin-filteringmass00, Hoeft:06, OkamotoGT08}.

The number of Milky Way and M31 satellites provides an indirect test
of galaxy formation and the importance of positive and negative
feedback in the early universe. This test, although the uncertainties
are large, supports the idea that a fraction of the new ultra-faint
dwarfs are fossils.  The good agreement of the SDSS and new M31
ultra-faint dwarf properties with predictions of our simulations
(RG05, GK06, Bovill \& Ricotti 2009) does not prove the primordial
origin of the new ultra-faint dwarfs, but it supports this
possibility.

More theoretical work and more observational data are needed to prove
that some dwarfs in the Local Group are true fossils of the first
galaxies. Future theoretical work should focus on improving the
accuracy of predictions on the properties of dwarf galaxies formed
before reionization and their evolution to the present day. Modeling
the evolution of the baryonic component after reionization in dwarf
satellites and in the Milky Way -- Andromeda system may be necessary
to make robust predictions. More observational data will certainly be
available in the near future. A large number of surveys, both at
optical and radio wavelengths will be online in the near future (\eg,
{\it Pan-STARRS, LSST, ALMA, EVLA, JWST, SKA} to mention a
few). Different survey strategies may be used to find and characterize
fossil dwarf galaxies. A deep pencil beam survey would be useful to
find the faintest dwarf satellites of the Milky Way and determine more
precisely their Galactocentric distribution. A shallower all sky
survey could be used to quantify the degree of anisotropy in the
distribution of satellites around the Milky Way.

The star formation history of the dwarf galaxies is not strongly
discriminatory because fossil galaxies may have a late phase of gas
accretion and star formation during the last $9-10$~Gyrs
\citep{Ricotti:09}. The distinction between fossils and non-fossil
galaxies may be quite elusive but it is nevertheless important to
understand galaxy formation and feedback in the early universe.
Arguments based on counting the number of dwarfs in the Local universe
are among the more solid arguments that could be used to prove the
existence of fossil galaxies (see Table~\ref{tab:count}). 

Future tests may be provided by deep surveys looking for ultra-faint
galaxies in the local voids or looking for gas in dark galaxies (\ie,
dark halos that have been able to accrete gas from the IGM at
$z<1-2$). Ultra-faint dwarfs should be present in the voids if dwarf
galaxies formed in large numbers before reionization (Bovill \&
Ricotti, in preparation). If pre-reionization dwarfs never formed due
to dominant negative feedback in the early universe, it is possible
that a faint (in H$\alpha$ and 21~cm emission) population of dark
galaxies exists in the outer parts of the Local Group. Hence, another
way to detect fossil galaxies in the outer parts of the Milky Way or
outside the super-galactic plane would be to search for neutral or
ionized gas that they may have accreted from the IGM. Future radio
telescopes (\eg, {\it ALMA, EVLA, SKA}) may be able to detect neutral
hydrogen in dark galaxies or in ultra-faint dwarfs. Ionized gas in the
outer parts of dark halos may be observed in absorption along the line
of sight of distant quasars (for instance in \mbox{\ion{O}{6}} or
\mbox{\ion{O}{4}} with {\it COS} on the {\it HST}). However, the
probability that a line of sight toward a quasar intersects the
ionized gas collected from the IGM by dark or fossil galaxies might be
small. Additional theoretical work is required to address these
issues.

\bibliographystyle{apj} 
\bibliography{./2ndyear,/Users/ricotti/Latex/TeX/archive}

\end{document}